%% file: main.tex
\documentclass[submission,copyright,creativecommons]{eptcs}
\providecommand{\event}{FOCLASA 2014} 

\usepackage{subfigure}
\usepackage{amssymb,amsmath}
\usepackage{graphicx}
\usepackage{array}
\usepackage{stmaryrd}
\usepackage{mathtools}

\usepackage{tikzexternal}      


\input{commands}

\newtheorem{proposition}{Proposition}[section]

\newtheorem{definition}{Definition}[section]

\def\titlerunning{Handshaking Protocol for Distributed Implementation of Reo}
\def\authorrunning{N. Kokash}

\begin{document}

\title{Handshaking Protocol for Distributed Implementation of Reo}

\author{N. Kokash
\institute{
  Niels Bohrweg 1, 2333 CA Leiden, The Netherlands
}
\email{natallia.kokash@gmail.com}
}

\maketitle

\begin{abstract}
Reo, an exogenous channel-based coordination language, is a model for service coordination wherein services communicate through connectors formed by joining binary communication channels. In order to establish transactional communication among services as prescribed by connector semantics, distributed ports exchange ``handshaking'' messages signalling which parties are ready to provide or consume data. In this paper, we present a formal implementation model for distributed Reo with communication delays and outline ideas for its proof of correctness. To reason about Reo implementation formally, we introduce Timed Action Constraint Automata (TACA) and explain how to compare TACA with existing automata-based semantics for Reo. We use TACA to describe ``handshaking'' behavior of Reo modeling primitives and argue that in any distributed circuit remote Reo nodes and channels exposing such behavior commit to perform transitions envisaged by the network semantics.
\end{abstract}

\setlength{\subfigcapskip}{0.1cm}
\setlength{\abovecaptionskip}{0cm}
\setlength{\belowcaptionskip}{0cm}
\setlength{\textfloatsep}{10pt plus 1.0pt minus 2.0pt}

\input{introduction}

\input{reo}

\input{semantics}

\input{motivation}

\input{routing}
\input{coordination}
\input{discussion}

\input{conclusions}

\bibliographystyle{eptcs}
\bibliography{main}

\end{document}

%% file: commands.tex
\newcommand{\mergerNode}{\tikz{
	    \node[inner sep=0,label=right:$C$] (D1) {};
	    \node[point,left of=D1,node distance=4mm] (C1) {};
	    \node[inner sep=0,label=left:$A$,above left of=C1,node
            distance=4mm] (B1) {};
	    \node[inner sep=0,label=left:$B$,below left of=C1,node
            distance=4mm] (A1) {};
	    \draw[sync] (C1) to (D1);
	    \draw[channel] (A1) to (C1);
	    \draw[channel] (B1) to (C1); }}

\newcommand{\replicatorNode}{\tikz{
	    \node[inner sep=0,label=left:$A$] (D1) {};
	    \node[point,right of=D1,node distance=4mm] (C1) {};
	    \node[inner sep=0,label=right:$B$,above right of=C1,node
          distance=4mm] (B1) {};
	    \node[inner sep=0,label=right:$C$,below right of=C1,node
          distance=4mm] (A1) {};
	    \draw[channel] (D1) to (C1);
	    \draw[sync] (C1) to (A1);
	    \draw[sync] (C1) to (B1); }}

\newcommand{\Reo}{\textsf{Reo}}
\renewcommand{\Reo}{Reo}

\newcommand{\A}{\ensuremath{\mathcal{A}}}

\newcommand{\N}{\ensuremath{\mathcal{N}}}

\newcommand{\R}{\ensuremath{\mathcal{R}}}
\newcommand{\calP}{\ensuremath{\mathcal{P}}}

\newcommand{\calT}{\ensuremath{\mathcal{T}}}
\newcommand{\M}{\ensuremath{\mathcal{M}}}

\newcommand{\channel}[1]{\ensuremath{\mathsf{#1}}}

\newcommand{\AsyncDrain}{\channel{AsyncDrain}}
\newcommand{\FIFO}{\channel{FIFO}}
\newcommand{\Filter}{\channel{Filter}}

\newcommand{\LossySync}{\channel{LossySync}}

\newcommand{\Sync}{\channel{Sync}}
\newcommand{\SyncDrain}{\channel{SyncDrain}}
\newcommand{\Transform}{\channel{Transform}}

\def\lparal{\mathbin{\setbox0=\hbox{$\|$}%
  \dimen0=\dp0 \advance\dimen0 -1.5pt \dp0=\dimen0%
  \underline{\kern-1.5pt\box0\kern1.5pt}}}

\newcommand{\trace}[3]{\ensuremath{#1\overset{#2}{\Longrightarrow}#3}}
\newcommand{\longtrace}[3]{\ensuremath{#1\overset{#2}{\xRightarrow{\hspace*{2cm}}}#3}}
\newcommand{\transition}[3]{\ensuremath{#1\overset{#2}{\xrightarrow{\hspace*{1cm}}}#3}}
\newcommand{\longtransition}[3]{\ensuremath{#1\overset{#2}{\xrightarrow{\hspace*{3cm}}}#3}}

\newcommand{\transone}[3]{\ensuremath{#1\overset{#2}{\xrightarrow{\hspace*{1cm}}_1}#3}}

\newcommand{\hide}{\ensuremath{\mathord{\backslash}}}

\newcommand{\joingamma}{\mathbin{\bowtie_{\mkern1mu \gamma}}}
\newcommand{\joinomega}{\mathbin{\bowtie_{\mkern1mu \omega}}}
\newcommand{\la}{\langle}

\newcommand{\syncgamma}{\mathbin{|_{\gamma}}}
\newcommand{\syncomega}{\mathbin{|_{\omega}}}
\newcommand{\ra}{\rangle}

%% file: introduction.tex
\section{Introduction}
\label{sect:introduction}
Service-oriented systems (SOS) are composed of autonomous services deployed on remote machines and accessed through the network.
Reo coordination language~\cite{Arb04:mscs} is an extensible notation for compositional modeling and execution of SOS.
Services that have no prior knowledge about each other communicate through channel connectors which guarantee that each participant, service or client, receives right data at the right time. Each channel is a binary function that imposes synchronization and data constraints on input and output messages. Channels can be composed to realize complex behavioral protocols, including multi-party synchronous rendezvous. This approach enables models that are both concise and compositional, but it also makes operational semantics for Reo non-trivial.

The most basic semantic model for Reo is constraint automata~(CA)~\cite{BSA+06}. States or locations in CA represent configurations of data stored in the buffers of Reo networks, while transition labels are composed of (i) sets of channel ends where dataflow is observed simultaneously, and (ii) data constraints necessary to trigger such transitions. The CA for a Reo connector can be computed as a product of the CA for its parts (sub-connectors or channels). CA is the theoretical basis for validation and verification tools for Reo, which are integrated in a framework known as the Extensible Coordination Tools (ECT)\footnote{\url{http://reo.project.cwi.nl/}}.

The Quality of Service (QoS) of a SOS depends on the quality of its components, efficiency of the ``glue code'' that coordinates individual services, and quality of the communication network. To evaluate the QoS of SOS coordinated by Reo, we need to estimate time to deliver input messages supplied by services to the input ports of the circuit to their consumers - services listening to the output ports of the circuit.
The early semantic models for quantitative Reo~\cite{ACS+07,ACM+09} assumed that delays in channels do not affect the operational semantics of Reo. This assumption is not realistic and limits the degree of concurrency in the presence of transactions with different durations. Hence, a more refined semantic model for Reo~\cite{KCA10} was introduced to solve this problem.

Distributed Reo ports exchange technical messages signalling the readiness of coordinated services to provide or consume data.
It has been recognized that the implementation of Reo should be distributed to avoid performance bottlenecks~\cite{PCV+12}. The existing semantic models for Reo focus on the description of the observable dataflow and are not suitable for the implementation and QoS evaluation.
In this paper, we present a formal coordination protocol that can serve as foundational basis for distributed time-aware Reo implementation. One of the main issues is to decide which requests are considered simultaneous and should synchronize on each execution cycle. In our approach, we propose to use a timeout which each node in a network should wait for to acquire the information about pending requests on remote ports and decide which transition to fire. The timeout is chosen to guarantee that the information about the communication requests on boundary ports is propagated through the circuit.

The remainder of this paper is organized as follows. In Section~\ref{sect:reo}, we explain the basics of Reo. In Section~\ref{sect:automataModel},
we describe a semantic model for Reo used and extended in this paper. In Section~\ref{sect:motivation}, we explain the objectives of our work. In Section~\ref{sect:routing}, we present implementation semantics for basic types of Reo nodes and channels. In Section~\ref{sect:proof}, we explore the properties of our approach. Section~\ref{sect:discussion} overviews related work. Finally, Section~\ref{sect:conclusions} concludes the paper and outlines future work.

%% file: reo.tex
\section{Reo Coordination Language}
\label{sect:reo}
\newcommand{\T}[2]{{\ensuremath{\begin{array}{c} \{#1\} \\ #2 \end{array}}}}
\newcommand{\U}[2]{{\ensuremath{\{#1\} \; #2}}}
\begin{figure}[t]
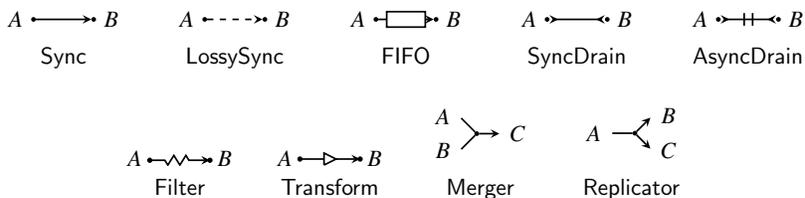

 \centering
 \scalebox{0.8}{
 \begin{tabular}{c}
     \begin{tabular}{c@{\qquad}c@{\qquad}c@{\qquad}c@{\qquad}c}
         \tikz{
    	   \node[point,label=left:$A$] (A) {};
    	   \node[point,right of=A,label=right:$B$] (B) {};
      	     \draw[sync] (A) -- (B); } &
         \tikz{
    	   \node[point,label=left:$A$] (A) {};
    	   \node[point,right of=A,label=right:$B$] (B) {};
      	     \draw[lossysync] (A) -- (B); } &
         \tikz{
            \node[point,label=left:$A$] (A) {};
            \node[point,right of=A,label=right:$B$] (B) {};
            \draw[fifo] (A) -- (B); } &
        \tikz{
            \node[point,label=left:$A$] (A) {};
            \node[point,right of=A,label=right:$B$] (B) {};
            \draw[syncdrain] (A) -- (B); } &
         \tikz{
            \node[point,label=left:$A$] (A) {};
            \node[point,right of=A,label=right:$B$] (B) {};
            \draw[asyncdrain] (A) -- (B); }
        \\
        \Sync & \LossySync & \FIFO & \SyncDrain & \AsyncDrain
      \end{tabular}
  \\
  \\
      \begin{tabular}{c@{\qquad}c@{\qquad}c@{\qquad}c}
          $A\;$\tikz{
            \filldraw[black] (0,0) circle (1pt) (1,0) circle (1pt);
            \draw[sync] (0,0) -- (0.25,0) -- (0.3, -0.08) -- (0.4, 0.08)
            -- (0.5, -0.08) -- (0.6,0.08) --(0.65, 0) -- (0.65,0) -- (1,0); }$\;B$ &
          $A\;$\tikz{
             \filldraw[black] (0,0) circle (1pt) (1,0) circle (1pt);
             \draw[channel,-] (0,0) -- (0.4,0);
             \draw[channel,-] (0.4, -0.1) -- (0.4, 0.1) -- (0.6, 0) --(0.4, -0.1);
             \draw[sync](0.6,0) -- (1,0); }$\;B$ &
          {\mergerNode} & {\replicatorNode}
        \\
        \Filter & \Transform & \channel{Merger} & \channel{Replicator}
      \end{tabular}
  \end{tabular}
  }
\caption{Graphical representation of basic Reo channels and nodes}
\label{fig:basicChannels}
\end{figure}
\noindent
{\Reo} is a coordination language in which components and services are coordinated exogenously by channel-based connectors~\cite{Arb04:mscs}. Connectors are graphs where the edges are user-defined communication channels and the nodes adhere to fixed routing rules.
Channels in {\Reo} are entities that have exactly two ends, also referred to as ports, which can be either \emph{source} or
\emph{sink} ends. Source ends accept data into, and sink ends dispense data out of their channel.

Although channels can be defined by users, a set
of basic Reo channels (see Figure~\ref{fig:basicChannels}) suffices to implement most common workflow patterns~\cite{AKS08}.
Among these channels are (i) the {\Sync} channel, which is a directed channel that accepts a data item through its source end if it can instantly dispense it through its sink end; (ii) the {\LossySync} channel, which always accepts a data item through its source end and tries to instantly dispense it through the sink end. If this is not possible, the data item is lost; (iii) the {\SyncDrain} channel, which is a channel with two source ends that accept data simultaneously and loses them subsequently; (iv) the {\AsyncDrain} channel, which accepts data items only through one of its two source channel ends at a moment in time and loses it; and (v) the {\FIFO} channel, which is an asynchronous channel with a buffer of capacity one. For data manipulation, Reo introduces the {\Filter} channel. which always accepts a data item at its source end and synchronously passes or loses it depending on whether or not the data item matches a certain predefined pattern or data constraint, and the {\Transform} channel, which applies a user-defined function to the data item at its source end and synchronously yields the result at its sink end.

Channels can be joined together using nodes. A node can be a \emph{source}, a \emph{sink} or a \emph{mixed} node. Source and sink nodes form the \emph{boundary} nodes of a connector to enable the  interaction with its environment. Source nodes act as synchronous \emph{replicators}, and sink nodes as non-deterministic \emph{mergers}. A mixed node combines these two behaviors by atomically consuming a data item from one of its sink ends at the time and replicating it to all of its source ends. Often two other nodes, \emph{route} and \emph{join}, are used to model non-deterministic routing and synchronization of flow, respectively. The router can be constructed from basic Reo channels while the \emph{join} node is a shorthand notation for a component that forms a tuple from data items received from several channel sink ports.

Channels can also differ at the level of their QoS. In quantitative Reo~\cite{ACS+07}, channels are characterized by a set of associated QoS parameters such as communication delays or cost. We recognize two types of communication delays: \emph{handshaking delay}, or time to decide whether the connector can satisfy the I/O request on its ends, and \emph{data transfer delay}, or the time needed to transfer the data accepted by the circuit.

%% file: semantics.tex
\section{Semantic models for Reo}
\label{sect:automataModel}
The semantics of any Reo connector can be better understood in terms of a specific semantic model and its appropriate translation into that model.

The most basic semantic model for Reo is constraint automata~(CA)~\cite{BSA+06}. Transitions in CA are
labeled with sets of ports that fire synchronously and data constraints on these ports. For example, a CA for the {\Sync} channel with port ends $A$ and $B$ contains one state ($s_0$) and one transition $\transition{s_0} {d_A = d_B, \{A,B\}}{s_0}.$ The {\FIFO} channel with ports $A$ and $B$ is described by a CA with two states corresponding to an empty buffer ($s_0$) and a full buffer ($s_1$), and transitions $\transition{s_0} {d = d_A, \{A\}}{s_1}$ and
$\transition{s_1} {d_B = d, \{B\}}{s_0}.$
The behavior of any Reo circuit can be computed using the product of CA of its basic channels. The \emph{hiding operator} is introduced to abstract from unnecessary details such as dataflow on the internal ports~\cite{BSA+06}.

Timed constrained automata~(TCA)~\cite{ABB+07} represent CA with clock assignments and timing constraints. 

The semantic models for Reo have been extended to compositionally compute QoS~\cite{ACS+07,ACM+09}, including communication delays in the circuit. It was assumed that delays do not affect operational semantics of the circuit and QoS labels were added to the transitions of basic CA. However, an example in~\cite{KCA10} shows that such approach limits concurrency in the circuit. Action constraint automata~(ACA) were introduced to overcome this issue~\cite{KCA10}. ACA distinguish several kinds of actions triggered on channel ports to signal the state changes of the channel:

\begin{definition}[ACA~\cite{KCA10}]
  \label{def:dca}
  An action constraint automaton $\A=(S, \N, \rightarrow, s_0)$ consists of a set
  of states~$S$, a set of action names~$\N$ derived from a set of port names $\M$ and a set of admissible action types $\calT$,
  a transition relation $\mathord{\rightarrow} \subseteq S \times 2^{\N} \times DC \times
  S$, where $DC$ is the set of data constraints over a finite data
  domain $\mathit{Data}$, and an initial state $s_0 \in S$.
\end{definition}

An ACA model proposed in~\cite{KCA10} uses the set of action types $\calT_1 = \{b, u\},$ where $b$ stands for the `block' and $u$ stands for the `unblock' actions to model synchronous channels with data transfer delays: when a channel is blocked, it does not accept new I/O requests. ACA is the generalization of CA, and CA can be seen as ACA with one action type: data flow through a Reo node.


To reason about the correctness of ACA as semantic models for Reo, we introduce a \emph{refinement} relation for ACA with various action types (in a specific case, for ACA and CA as the latter is equivalent to the ACA with one action: observation of dataflow on Reo ports). For simplicity we omit data constraints in the rest of this paper and focus on ACA synchronization constraints. First, we need to be able to say whether automata with various observable actions describe the same Reo circuit or not. Thus, we introduce an action renaming operator to unify sets of action names, show its compositionality and then define a weak bisimulation relation to compare ACA with renamed actions. Proofs for the following properties can be found in~\cite{Kok14}.

\begin{definition}[Action renaming]
For any $\A = (S, \N, \rightarrow, s_0),$ let $\rho(\A, R)$ be an action renaming operator where $R$ is a set of renamings in the form $x \rightarrow y, \, x \in H \subseteq \N, \, y = \varrho(x), \, \varrho: H \rightarrow H' \subseteq \N'.$
$\rho(\A, R) = (S, \N', \rightarrow', s_0)$ is an ACA such that
for any $(s,N, s') \in \rightarrow$ there exists $(s, N \setminus H \cup \varrho[N \cap H], s_0') \in \rightarrow'.$
\end{definition}

\begin{proposition}[Commutativity of renaming and hiding]
    $\rho(\texttt{hide}(\A, K), R) = \texttt{hide}(\rho(\A, R), (K \setminus H) \cup \varrho[K \cap H]).$
    \label{prop:renaming-hiding}
\end{proposition}

For $H \cap K = \emptyset$ it holds that $\rho(\texttt{hide}(\A, K), R) = \texttt{hide}(\rho(\A, R), K).$

As in~\cite{KKV10}, we use the port synchronization function~$\gamma$ as follows: we
write~$\N'_1$ for $\N_1 \hide \gamma_1 [ \N ]$ and~$\N'_2$ for $\N_2
\hide \gamma_2 [ \N ]$.  If, for subsets $N_1 \subseteq \N_1$, $N_2
\subseteq \N_2$, it holds that $\gamma_1^{-1}[ N_1 ] = \gamma_2^{-1} [
  N_2 ]$ we write $N_1 \syncgamma N_2 = ( N_1 \cap \N'_1 ) \cup \gamma_1^{-1}[ N_1 ]\cup ( N_2 \cap \N'_2 ).$
Hence, $N_1 \syncgamma N_2$ is the union $N_1 \cup N_2$ but with the parts of $N_1$ and~$N_2$ that
are identified via $\gamma_1$ and~$\gamma_2$ replaced by the shared names $\gamma_1^{-1}[N_1] = \gamma_2^{-1}[N_2]$.
The following proposition states that the action renaming is compositional provided that in the product of ACA we rename the set of synchronized actions that is obtained from the sets of renamed actions in the original automaton:
\begin{proposition}[Compositionality of action renaming]
  \label{prop:renaming}
  Let $\A_1 = (S_1, \N_1, \rightarrow_{\A_1}, s_0^1)$ and $\A_2 = (S_2, \N_2,$ $ \rightarrow_{\A_2}, s_0^2)$ be two ACA with disjoint sets of action names, $\N_1 \cap \N_2 = \emptyset.$
   Let also $\gamma: \N \rightarrow \N_1 \times \N_2$ be an action synchronization function defined as $\gamma(n) = (\gamma_1(n), \gamma_2(n)),$ where
   $\gamma_1: \N \rightarrow \N_1, \, \gamma_2 : \N \rightarrow \N_2$ is a set of injective functions that map action names from the new set $\N$
   into action names from the initial sets $\N_1$ and $\N_2,$
   $\N \cap (\N_1 \cup \N_2) = \emptyset.$
   Given sets of renamings $R_1: \{x \rightarrow y = \varrho_1(x), \, \varrho_1: H_1 \rightarrow L_1, H_1 \in \N_1\}$ and
   $R_2: \{x \rightarrow y = \varrho_2(x), \, \varrho_2: H_2 \rightarrow L_2, H_2 \in \N_2\},$ for $\A_1$ and $\A_2,$ respectively, and a set of renamings
   $R = \{x \rightarrow y = \varrho(x)\}$ for their product, where
   \[
   \varrho(x) =
   \begin{cases}
      \varrho_1(x) & x \in H_1 \subseteq H_1 \syncgamma H_2 \\
      \varrho_2(x) & x \in H_2 \subseteq H_1 \syncgamma H_2 \\
      f(\varrho_1(x_1), \varrho_2(x_2)) & x = \gamma_1^{-1}(x_1) = \gamma_2^{-1}(x_2), \, x_1 \in H_1, \, x_2 \in H_2
   \end{cases}
    \]
    it holds that $$\rho(\A_1 \joingamma \A_2, R) = \rho(\A_1, R_1) \joinomega \rho(\A_2, R_2).$$
    Here $\omega: \M \rightarrow \M_1 \times \M_2,$ is an action synchronization function for the renamed actions defined as
    $\omega(n) = (\omega_1(n), \omega_2(n)),$ $\omega_1: \M  \rightarrow \M_1, \, \omega_2: \M \rightarrow \M_2, \,$
    \[
    \begin{array}{c}
        \M = \varrho[H_1 \syncgamma H_2] \cup \N \setminus (H_1 \syncgamma H_2),\\
        \M_1 = \varrho[H_1] \cup \N_1 \setminus H_1, \quad \M_2 = \varrho[H_2] \cup \N_2 \setminus H_2,\\
        \gamma_1(n) = n_1 \wedge \gamma_2(n) = n_2 \mbox{ iff } \omega_1(\varrho(n)) = \varrho(n_1) \wedge \omega_2(\varrho(n)) = \varrho(n_2).
    \end{array}
    \]
\end{proposition}

Note that hiding can be seen as renaming to unobservable action $\tau.$ Hence, it is compositional under the same conditions.

We define traces for ACA in a usual way: a finite or infinite sequence of transitions $$r = \transition{s}{N_0}{s_1} \transition{}{N_1}{s_2} \transition{}{N_2}{s_3} \, . . .$$ is an $s$-trace in $ACA.$
Let $S^*$ be the set of all finite sequences over a set $S.$ Given finite sequences $\sigma_1$ and $\sigma_2$, we denote their concatenation $\sigma_1 \cdot \sigma_2$. If for some ACA there exists an $s$-trace
$$\transition{s}{N_1}{} \transition{s_1}{\emptyset}{} \transition{s_2}{\emptyset}{}
\transition{s_3}{N_2}{} \transition{s_4}{\emptyset}{} \transition{s_5}{N_3}{}
\transition{s_6}{\emptyset}{s'},$$
where $N_1,N_2,N_3 \subseteq \N$ are sets of actions representing ACA labels, we write
$\longtransition{s}{N_1 \cdot \emptyset \cdot \emptyset \cdot N_2 \cdot \emptyset \cdot N_3 \cdot \emptyset}{s'},$
or
$\longtrace{s}{N_1 \cdot N_2 \cdot N_3}{s'}$ for the empty action set abstracted traces.

\begin{definition}[Weak bisimulation]
    \label{def:weak_bisim_ACA}
     Let $\A = (S, \N, \rightarrow, s_0)$ be an ACA. A weak bisimulation on $\A$ is an equivalence $\Theta$ on $S$ such that for all $(s_1, s_2) \in \Theta$
     if $\trace{s_1}{N}{p_1}$ then $\trace{s_2}{N}{p_2}$ for some $(p_1, p_2) \in \Theta.$
\end{definition}
States $s_1$ and $s_2$ are called weakly bisimilar iff there exists weak bisimulation $\R$ such that $(s_1, s_2) \in \R.$

\begin{definition}[Action refinement]
  \label{def:refinement}
   Let $\A = (S, \N, \rightarrow_{\A}, s_0)$ and $\mathcal{B} = (Q, \M, \rightarrow_{\mathcal{B}}, q_0)$ be two ACA.
   We say that $\A$ is an action refinement of $\mathcal{B}$, written as $\mathcal{B} \preccurlyeq \A$, iff
   there exist a set $K \subseteq \N$ and a set of renamings $R = \{x \rightarrow y = \varrho(x), \, \varrho : \N \setminus K \rightarrow \M \}$ such that $\mathcal{B}$ and $\rho(\texttt{hide}(\A, K), R) = (S, \M, \rightarrow', s_0)$ are weakly bisimilar.
\end{definition}

\begin{proposition}[Compositionality of action refinement]
 Let $\A_1 = (S_1, \N_1, \rightarrow_{\A_1}, s_0^1)$ and $\mathcal{B}_1 = (Q_1,$ $\M_1, \rightarrow_{\mathcal{B}_1}, q_0^1)$ be two
 ACA such that $\mathcal{B}_1 \preccurlyeq A_1$ with
 a set of hidden actions $K_1 \subseteq \N_1$
 and a set of renamings $R_1: \{x \rightarrow y = \varrho_1(x), \, \varrho_1: \N_1 \setminus K_1 \rightarrow \M_1\}.$ Let $\A_2 = (S_2, \N_2, \rightarrow_{\A_2}, s_0^2)$ and $\mathcal{B}_2=(Q_2, \M_2, \rightarrow_{\mathcal{B}_2}, q_0^2$ be two ACA such that $\mathcal{B}_2 \preccurlyeq \A_2$ with a set of hidden actions $K_2 \subseteq \N_2$ and a set of renaming
 $R_2: \{x \rightarrow y = \varrho_2(x), \, \varrho_2: \N_2 \setminus K_2 \rightarrow \M_2\}.$
 Assume also that $\N_1 \cap \N_2 = \M_1 \cap \M_2 = \emptyset.$

 The $\gamma$-synchronous product of automata $\A_1$ and $\A_2$ is the action refinement of the $\omega$-synchronous product of automata $\mathcal{B}_1$ and $\mathcal{B}_2,$ i.e., $$\mathcal{B}_1 \joinomega \mathcal{B}_2 \preccurlyeq \A_1 \joingamma \A_2$$
 with a set of hidden actions $K = K_1 \syncgamma K_2$ and a set of renamings
 $R = \{x \rightarrow y = \varrho(x): (\N_1 \setminus K_1) \syncgamma (\N_2 \setminus K_2)\} \rightarrow \M_1 \syncomega \M_2$ where $\varrho(x)$ is defined as in Prop.~\ref{prop:renaming},
 $\gamma = (\gamma_1, \gamma_2), \, \gamma_1: \N \rightarrow \N_1, \, \gamma_2: \N \rightarrow \N_2$ and
 $\omega = (\omega_1, \omega_2), \, \omega_1: \M \rightarrow \M_1, \, \omega_2: \M \rightarrow \M_2$
 are action synchronization functions such that
     \[
    \begin{array}{c}
     \N \cap (\N_1 \cup \N_2) = \emptyset, \quad \M \cap (\M_1 \cup \M_2) = \emptyset, \mbox{ and }\\
    \gamma_1(n) = n_1 \wedge \gamma_2(n) = n_2 \mbox{ iff } \omega_1(\varrho(n)) = \varrho(n_1) \wedge \omega_2(\varrho(n)) = \varrho(n_2).
    \end{array}
    \]
\end{proposition}

%% file: motivation.tex
\section{Semantic model for Reo implementation}
\label{sect:motivation}
Existing semantic models for Reo describe the coordination behavior of Reo circuits but do not show how to achieve it. We refer to the process of exchanging messages among remote Reo nodes in order to establish whether they are ready to accept or provide data as handshaking protocol: before transmitting service data, Reo nodes notify each other about their internal states. It is not clear how Reo nodes and channels should behave to determine which transitions are enabled and agree to perform one transition from the set of enabled ones. Specifying such behavior is important for the generation of executable coordination code~\cite{PCV+12}.

\begin{figure}[h]
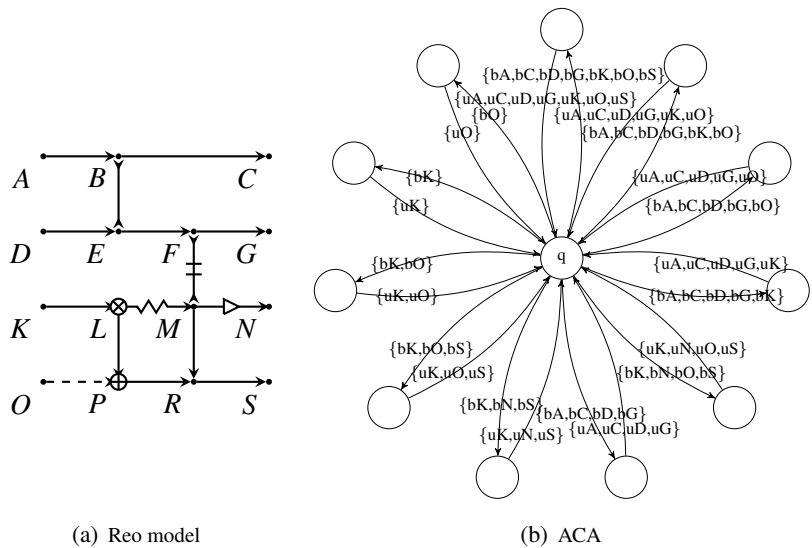

\centering
    \subfigure[\scriptsize{Reo model}]{
        \raisebox{1cm}{
            \scalebox{1.0}{
                \tikz{
                    \node[point,label=below left:$A$] at (0, 2) (A) {};
                    \node[point,label=below left:$B$] at (1, 2) (B) {};
                    \node[point,label=below left:$C$] at (3, 2) (C) {};
                    \node[point,label=below left:$D$] at (0, 1) (D) {};
                    \node[point,label=below left:$E$] at (1, 1) (E) {};
                    \node[point,label=below left:$F$] at (2, 1) (F) {};
                    \node[point,label=below left:$G$] at (3, 1) (G) {};
                    \node[point,label=below left:$K$] at (0, 0) (K) {};
                    \node[point,label=below left:$L$] at (1, 0) (L) {};
                    \node[point,label=below left:$M$] at (2, 0) (M) {};
                    \node[point,label=below left:$N$] at (3, 0) (N) {};
                    \node[point,label=below left:$O$] at (0, -1) (O) {};
                    \node[point,label=below left:$P$] at (1, -1) (P) {};
                    \node[point,label=below left:$R$] at (2, -1) (R) {};
                    \node[point,label=below left:$S$] at (3, -1) (S) {};
                    \draw[sync]  (A) to node {} (B);
                    \draw[sync]  (B) to node {} (C);
                    \draw[syncdrain]  (B) to node {} (E);
                    \draw[sync]  (D) to node {} (E);
                    \draw[sync]  (E) to node {} (F);
                    \draw[sync]  (F) to node {} (G);
                    \draw[sync]  (K) to node {} (L);
                    \draw[sync] (1,0) -- (1.25,0) -- (1.3, -0.08) -- (1.4, 0.08)
                    -- (1.5, -0.08) -- (1.6,0.08) --(1.65, 0) -- (1.65,0) -- (2,0);
                    \draw[channel,-] (2,0) -- (2.4,0);
                    \draw[channel,-] (2.4, -0.1) -- (2.4, 0.1) -- (2.6, 0) --(2.4, -0.1);
                    \draw[sync](2.6,0) -- (3,0);
                    \draw[asyncdrain]  (F) to node {} (M);
                    \draw[sync]  (L) to node {} (P);
                    \draw[sync]  (M) to node {} (R);
                    \draw[lossysync]  (O) to node {} (P);
                    \draw[sync]  (P) to node {} (R);
                    \draw[sync]  (R) to node {} (S);
                    \draw [-, thick, fill=white] (1, 0) circle (3pt);
                    \draw [-, thick](0.93, -0.07) -- (1.07, 0.07);
                    \draw [-, thick](0.93, 0.07) -- (1.07, -0.07);
                    \draw [-, thick, fill=white] (1,-1) circle (3pt);
                    \draw [-, thick](0.9, -1) -- (1.1,-1);
                    \draw [-, thick](1, -0.9) -- (1,-1.1);
                }
            }
        }
    }
    \subfigure[\scriptsize{ACA}]{
        \scalebox{0.8}{
        \scriptsize
            \tikz{
               \node[state] (q) {q};
                \def\labels{q1,q2,q3,q4,q5,q6,q7,q8,q9,q10,q11}\def\dim{11}
                \foreach \n [count=\ni] in \labels {%
                  \node[state={\n}] at ({cos(90+\ni*(360/\dim))*3.8},{sin(90+\ni*(360/\dim))*3.8}) (\n) {};}
               \path[transition]
               (q)  edge [bend right=15] node[near end] { \{bO\} } (q1)
               (q1) edge [bend right=15] node[near start] { \{uO\} } (q)
               (q) edge [bend right=15] node[near end]  { \{bK\} } (q2)
               (q2) edge [bend right=15] node[near start] { \{uK\} } (q)
               (q) edge [bend right=15] node[near end] { \{bK,bO\} } (q3)
               (q3) edge [bend right=15] node[near start] { \{uK,uO\} } (q)
               (q) edge [bend right=15] node[near end]  { \{bK,bO,bS\} } (q4)
               (q4) edge [bend right=15] node[near start] { \{uK,uO,uS\} } (q)
               (q) edge [bend right=15] node[near end] { \{bK,bN,bS\} } (q5)
               (q5) edge [bend right=15] node[very near start] { \{uK,uN,uS\} } (q)
               (q) edge [bend right=15] node[near end]  { \{bA,bC,bD,bG\} } (q6)
               (q6) edge [bend right=15] node[very near start] { \{uA,uC,uD,uG\} } (q)
               (q) edge [bend right=15] node[near end]  { \{bK,bN,bO,bS\} } (q7)
               (q7) edge [bend right=15] node[near start] { \{uK,uN,uO,uS\} } (q)
               (q) edge [bend right=15] node[near end]  { \{bA,bC,bD,bG,bK\} } (q8)
               (q8) edge [bend right=15] node[near start] { \{uA,uC,uD,uG,uK\} } (q)
               (q) edge [bend right=15] node[near end]  { \{bA,bC,bD,bG,bO\} } (q9)
               (q9) edge [bend right=15] node[near start] { \{uA,uC,uD,uG,uO\} } (q)
               (q) edge [bend right=15] node[near end]  { \{bA,bC,bD,bG,bK,bO\} } (q10)
               (q10) edge [bend right=15] node[near start] { \{uA,uC,uD,uG,uK,uO\} } (q)
               (q) edge [bend right=15] node[very near end]  { \{bA,bC,bD,bG,bK,bO,bS\} } (q11)
               (q11) edge [bend right=15] node[near start] { \{uA,uC,uD,uG,uK,uO,uS\} } (q);
            }
        }
    }
\caption{Complex synchronous circuit}
\label{fig:example}
\end{figure}

In Reo, large synchronous regions can be constructed. By synchronous region we understand part of a circuit consisting of joint synchronous channels. Figure~\ref{fig:example} shows a synchronous Reo circuit with eight boundary nodes and its operational semantics in the form of ACA. In this circuit, 4 source nodes may provide data, 4 sink nodes may consume data, while 7 mixed internal nodes together with channels  coordinate data flow. Here $L$ is an exclusive \emph{route} node, $M$ is a \emph{join} node, and other nodes and channels behave as explained in Section~\ref{sect:reo}. Assume that a write request arrives on node $A$. $A$ can accept this request iff (i) $D$ has a pending write request, (ii) nodes $C$ and $G$ are ready to accept, and (iii) there is no data flow on port $M$ (required by the semantics of the {\AsyncDrain} channel). If nodes are deployed on remote machines, the node cannot decide whether to accept data until it receives messages from all parties it depends on. If several transitions are enabled (as shown by the ACA semantics, 11 transitions are possible in our example if all boundary nodes are ready to communicate), the implementation should non-deterministically choose one of them.

The goal of the handshaking protocol is to ensure that the internal choice made by Reo nodes locally lead to the correct implementation of the global observable behavior of the circuit. For example, if there are pending requests on nodes $K$, $O$ and $S$, but no request on node $N$, the exclusive router $L$ should transfer data to the node $P$ and the merge node $R$ should consume data from node $P$ because transition $\{K, L, O, P, R, S\}$ (corresponds to an observable transition $\{K, O, P\})$ is enabled and should not be excluded by a local choice of a node or a channel with non-deterministic behavior.

The existing implementations for Reo either (i) decompose a circuit to synchronous regions and deploy all nodes and
channels from such a region on a single machine~\cite{MLA08},
(ii) propagate technical messages to establish which nodes are ready to process data as if there were no
delays~\cite{PCV+12}. The first approach is not a fully distributed solution: if the whole network consists of a single synchronous region, the coordination is performed by a single machine.
The second approach implies no constraints on node deployment but it is not efficient for two reasons. Firstly, after an internal node receives a request message, it propagates it to the rest of the network. Meanwhile, if another message comes to the same node, it needs to be propagated again. Thus, internal nodes update their routing tables several times per execution cycle. Secondly, on each execution cycle, one node is elected to resolve non-deterministic choice and need to keep a list of enabled nodes for the whole synchronous region.

\begin{figure}
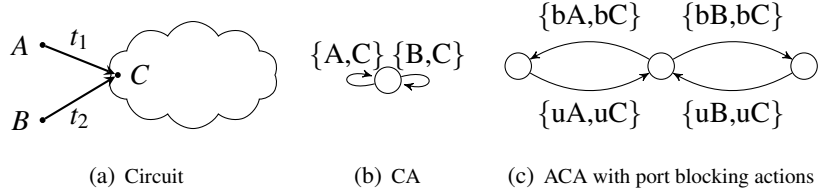

\centering
  \scalebox{1.0}{
    \subfigure[\scriptsize{Circuit}]{
    \tikz{
        \node[point,label=left:$A$] at (0, 0) (A) {};
        \node[point,label=left:$B$] at (0, -1) (B) {};
        \node[point,label=right:$C$] at (1, -0.4) (C) {};
        \draw[sync]  (A) to node[above] {$t_1$} (C);
        \draw[sync]  (B) to node[below] {$t_2$} (C);
        \node [cloud, draw,cloud puffs=10,cloud puff arc=120, aspect=2, inner ysep=1em] at (2, -0.4) () {};
    }}
    \subfigure[\scriptsize{CA}]{
    \raisebox{0.6cm}{
    \tikz{
       \node[state] (q) {};
       \path[transition]
       (q)  edge [loop left] node[above] { \{A,C\} } (q)
       (q) edge [loop right] node[above] { \{B,C\} } (q); }
    }
    }
    \subfigure[\scriptsize{ACA with port blocking actions}]{
    \tikz{
       \node[state] (q) {};
       \node[state,right of=q] (q1) {};
       \node[state,left of=q] (p1) {};
       \path[transition]
       (q)  edge [bend left] node[above] { \{bB,bC\} } (q1)
       (q1) edge [bend left] node[below] { \{uB,uC\} } (q)
       (q) edge [bend right] node[above]  { \{bA,bC\} } (p1)
       (p1) edge [bend right] node[below] { \{uA,uC\} } (q) ; }
    }
  }
\caption{Merge circuit and its semantics}
\label{fig:mergeNode}
\end{figure}

Consider a circuit with a merge fragment in Figure~\ref{fig:mergeNode}.
Assuming that both input ports $A$ and $B$ receive write requests simultaneously,
in the time-agnostic implementation~\cite{PCV+12}, the merge node $C$ propagates the
request that arrives first. I.e., if $t_1 < t_2$, the merge node assumes that only
transition $\{A,C\}$ is enabled. In time $t_2 - t_1$ it learns that $\{B, C\}$ is enabled as
well and sends a new technical message to notify other nodes in the synchronous region.
This causes undesired traffic and does not scale well. Furthermore, the non-deterministic
choice introduced by $C$ is not resolved locally, but delegated to an external node
(elected randomly, depending on its ID). Thus, the existing implementation is not
optimal in time-aware environment and relies on centralized resolution of non-determinism at each execution cycle to ensure absence of inconsistencies.

We need a time-aware sematic model for Reo that displays their internal state changes (e.g., from \emph{idle} to \emph{waiting for reply} to \emph{committed} and back to \emph{idle}). TCA~\cite{ABB+07} models circuit time delays while ACA~\cite{KCA10} allows multiple actions to be observed on node and channel ports. We combine these two models to describe our handshaking protocol.

Let $\mathcal{C}$ be a finite set of clocks and $v: \mathcal{C} \rightarrow \mathbb{R}_{\geq 0}$ be a clock assignment function as defined in~\cite{ABB+07}.
Let also $cc$ be a clock constraint for $C$, which is defined as a conjunction of atoms of the form $x \odot n$ where $x \in \mathcal{C}, \odot \in \{<, \leq, >, \geq, =\}$ and $n \in \mathbb{N}$. $CA(\mathcal{C})$ (or $CA$) denotes the set of all clock assignments and $CC(\mathcal{C})$ (or $CC$) the set of all clock constraints.

\begin{definition}[Timed ACA]
  \label{def:TACA}
  A Timed ACA (TACA) is a tuple $\A=(S, \mathcal{C}, \N, \rightarrow, s_0, ic),$ where
  $S$ is a finite set of states, $\mathcal{C}$ is a finite set of clocks,
  $\N$ is a set of action names derived from a set of port names $\M$ and a set of admissible action types $\calT$,
  $\mathord{\rightarrow} \subseteq S \times 2^{\N} \times DC \times CC \times 2^\mathcal{C} \times S$ is a transition relation such that
  $DC$ is the set of data constraints over a finite data domain $\mathit{Data}$,
  $s_0 \in S$ is an initial state,
  $ic: S \rightarrow CC$ is a function that assigns to any location $s$ an invariance condition $ic(s)$.
\end{definition}

Let an injective function $act: \M \times \calT \rightarrow {\N} $ define action names for each pair of a port name and an action type observed on the port as discussed in~\cite{KCA10}. 
The action synchronization function~$\gamma : \N \rightarrow \N_1 \times \N_2 $ is defined through
a pair of injective functions $\gamma_1 \colon \N \to \N_1$, $\gamma_2 \colon \N \to \N_2$ from a
new set of action names~$\N$ into $\N_1$ and~$\N_2$~\cite{KKV10}.

\begin{definition}[Product of TACA]
  For two TACA $\A_1 = (S_1, \mathcal{C}_1, \N_1, \rightarrow_1, s_0^1, ic_1)$ and~$\A_2 = (S_2, \mathcal{C}_2, \N_2,
  \rightarrow_2, s_0^2, ic_2)$ and the action
  synchronization function~$\gamma : \N \rightarrow \N_1 \times \N_2$ with $\gamma_1: \N \rightarrow \N_1$ and $\gamma_2 : \N \rightarrow \N_2$,
  the TACA $\A_1 \joingamma \A_2$,
  called the $\gamma$-synchronization product of $\A_1$
  and~$\A_2$, is given by $\A_1 \joingamma \A_2 = ( S_1 \times S_2, \mathcal{C}_1 \cup \mathcal{C}_2,
  \N'_1 \syncgamma \N'_2, {\to}, \la s_0^1 , s_0^2 \ra, ic(\la s_1, s_2 \ra))$ where  $ic(\la s_1, s_2 \ra) = \chi(ic(s_1)) \wedge \chi(ic(s_2)),$
  for any $s_1 \in S_1, s_2 \in S_2,$ $\chi: CC_1 \cup CC_2 \rightarrow CC$ is a clock constraint and location invariance condition update function, and the transition relation~$\to$ is determined by the following rules:
  \begin{equation}
    \label{eq:join-rule-1}
    \frac{ \
      \transone{s_1}{N_1,g_1,cc_1}{t_1} \quad
      N_1 \subseteq \N'_1 \ }{
      \transition{
        \langle s_1,s_2 \rangle}{N_1, g_1, \chi(cc_1)}{
        \langle t_1,s_2 \rangle}
    }
    \qquad
    \frac{ \
      \transone{s_2}{N_2,g_2,cc_2}{t_2} \quad
      N_2 \subseteq \N'_2 \ }{
      \transition{
        \langle s_1,s_2 \rangle}{N_2, g_2, \chi(cc_2)}{
        \langle s_1,t_2 \rangle}
    }
  \end{equation}
  and
  \begin{equation}
    \label{eq:join-rule-23}
    \frac{ \
      \transone{s_1}{N_1,g_1,cc_1}{t_1} \quad
      \transone{s_2}{N_2,g_2,cc_2}{t_2} \quad
      \gamma_1^{-1}(N_1) = \gamma_2^{-1}( N_2 ) \ }{
      \longtransition{
        \langle s_1,s_2 \rangle}{
        N_1 \syncgamma N_2, \gamma(g_1 \land g_2), \gamma(\chi(cc_1) \land \chi(cc_2))}{
        \langle t_1,t_2 \rangle}
    }
  \end{equation}
\end{definition}

Function $\chi$ is introduced to update invariance conditions and clock constraints in composed circuits.
In particular, this extension is needed to increase timeouts for composed networks. For example, if invariance conditions in $\A_1$ are in the form $x \leq T_1$ and the invariance conditions in $\A_2$ are in the form $x \leq T_2$, by setting $\chi(x \leq T_1) = x \leq T_1 + T_2,$ and $\chi(x \leq T_2) = x \leq T_1 + T_2,$ we extend timeout needed for the traversal of the composed network. Consequently, clock constraints in the form $x > T_1$ and $x > T_2$ are replaced with $x > T_1 + T_2.$

We define the hiding operator \texttt{hide}$(\A, K),$ where $K$ is a non-empty set of actions $K \subseteq \N,$ and a state-transition graph of TACA $G_{\A}$ analogously to the TCA~\cite{ABB+07}. The only distinction is that instead of the nodes set as in CA we deal with the action sets as in ACA.

To reason about time-agnostic Reo semantics, we need to abstract from time in TACA. We omit data constraints and focus on action synchronization constraints.
Let $q = \langle s, v\rangle$ be a state of a state-transition graph of TACA $G_{\A}$. We call a finite or
infinite sequence of transitions
$$r = \transition{q}{N_0, t_0}{q_1} \transition{}{N_1, t_1}{q_2} \transition{}{N_2, t_3}{q_3} \, . . .$$
a \emph{$q$-trace} in $G_{\A}.$
We say that a finite or infinite sequence of transitions
$$r = \transition{q}{N_0}{q_1} \transition{}{N_1}{q_2} \transition{}{N_2}{q_3} \, . . .$$ is an \emph{untimed $q$-trace} iff there exist $t_0, t_1, t_2,... \in \mathbb{R_{\geq}}$ such that $$r' = \transition{q}{N_0, t_0}{q_1} \transition{}{N_1, t_1}{q_2} \transition{}{N_2, t_2}{q_3} \, . . .$$
is a $q$-trace in $G_{\A}.$

Let $Q^*$ be the set of all finite sequences over a set $Q=\{\langle s, v \rangle \, | \,  s \in S \}.$ Given finite sequences $\sigma_1$ and $\sigma_2$, we denote their concatenation $\sigma_1 \cdot \sigma_2$. If for some TACA there exists an untimed trace $\longtransition{q}{N_1 \cdot \{\} \cdot \{\} \cdot N_2 \cdot \{\} \cdot N_3 \cdot \{\}}{p},$ where $ N_1,N_2,N_3 \in 2^\N$ are sets of actions representing TACA labels, we write $\longtrace{q}{N_1 \cdot N_2 \cdot N_3}{p}.$

%% file: routing.tex
\section{Reo handshaking protocol}
\label{sect:routing}
I/O requests arrive to the Reo source nodes from \emph{writers}, or to the sink nodes from \emph{readers}. Writers and readers for synchronous regions are either external components or buffered Reo channels. Once a pending request is detected, the boundary node initiates handshaking message exchange through channels that connect it with its neighbors, reporting its status and requesting to confirm the ability to accept or provide data. At this stage, channels work as simple communication links between adjacent nodes regardless of their semantics.

Three message propagation strategies are possible: (i) \emph{forward propagation}, when the handshaking is initiated by writers, readers remain passive and reply to the arrived messages; (ii) \emph{backward propagation}, when the communication is initiated by readers, writers are passive; and (iii) \emph{two-side propagation}, when both writers and readers can initiate the message exchange. Two-side propagation minimizes handshaking delay, but it is also more difficult to implement. In the remainder of this paper we consider forward propagation.

The handshaking behavior of Reo nodes depend on their type (input, output, simple internal, merge, replicate, route, or join) and (the number of) adjacent channels. Nodes can exchange three types of messages: (i) intention to write (\emph{write}), (ii) possibility to read (\emph{read}), and (iii) possibility to write (\emph{may\_write}). The third message type is needed for routers to obtain status of nodes in alternative branches without giving definite promise to write data. Together with three message types, four actions are recognized at each channel port: (i) send a message to an adjacent node, (ii) receive a message from an adjacent node, (iii) block (or commit) port, (iv) unblock the port.
\begin{figure}
 \centering
    \begin{tabular}{c}
    \subfigure[\scriptsize{Source node}]{
    \scalebox{0.7}{
     \begin{tikzpicture}
            \node[state] (q) {$s_0$};
            \node[state,right of=q] (q1) {$x<T$};
            \node[state,below of=q1] (q2) {};
            \node[state,below of=q] (commit) {$x<T$} ;
        \path[transition]
             (q)  edge node[below]  { \{$?wA_{out}$\},x:=0} (q1)
             (q1) edge[bend right] node[above] { \{\},$x>T$} (q)
             (q1) edge node[below]  { \{$!rA_{out}$\} } (q2)
             (q2) edge node[above]  { \{$bA$\},x:=0} (commit)
             (commit) edge node[below]  { \{$uA$\},$x>T$ } (q) ;
     \end{tikzpicture}
     \label{fig:TACA_source}
    }}
    \subfigure[\scriptsize{Sink node}]{
     \scalebox{0.7}{
    \begin{tikzpicture}
            \node[state] (q) {$s_0$};
            \node[state,right of=q] (q1) {};
            \node[state,below of=q1] (q2) {};
            \node[state,below of=q] (commit) {$x<T$};
            \node[state,left of=q] (p1) {};
            \node[state,below of=p1] (p2) {$x<T$};
        \path[transition]
             (q)  edge node[above]  { \{$!wA_{in}$\},x:=0} (q1)
             (q1) edge[bend right, dashed] node[above]  { \{\}} (q)
             (q1) edge node[below]  { \{$?rA_{in}$\}} (q2)
             (q2) edge node[below]  { \{$bA$\},x:=0} (commit)
             (commit) edge node  { \{$uA$\},$x>T$ } (q)
             (q)  edge node[above]  { \{$!mwA_{in}$\},x:=0} (p1)
             (p1) edge[bend left] node[above]  { \{\}} (q)
             (p1) edge node[above]  { \{$?rA_{in}$\} } (p2)
             (p2) edge node[above]  { \{\},$x>T$ } (q)
             (p2) edge node[near start]  { \{$!wA_{in}$\}} (q1);
     \end{tikzpicture}
     \label{fig:TACA_sink}
    }}
    \subfigure[\scriptsize{Mixed node}]{
     \scalebox{0.7}{
    \begin{tikzpicture}
            \node[state] (q) {$s_0$};
            \node[state,right of=q] (q1) {};
            \node[state,right of=q1] (q2) {$x<T$};
            \node[state,below of=q2] (q3) {};
            \node[state,below of=q1] (q4) {};
            \node[state,below of=q] (commit) {$x<T$};
            \node[state,left of=q] (p1) {};
            \node[state,left of=p1] (p2) {$x<T$};
            \node[state,below of=p2] (p3) {};
            \node[state,below of=p1] (p4) {$x<T$};
        \path[transition]
             (q)  edge node[above]  { \{$!wA_{in}$\} } (q1)
             (q1) edge node[above]  { \{$?wA_{out}$\},x:=0} (q2)
             (q2) edge[bend right] node[above] { \{\},$x>T$ } (q)
             (q2) edge node[above]  { \{$!rA_{out}$\} } (q3)
             (q3) edge node[above]  { \{$?rA_{in}$\}} (q4)
             (q4) edge node[above]  { \{$bA$\},x:=0} (commit)
             (commit) edge node  { \{$uA$\},$x>T$ } (q)
             (q)  edge node[above]  { \{$!mwA_{in}$\} } (p1)
             (p1) edge node[above]  { \{$?mwA_{out}$\}, x:=0} (p2)
             (p2) edge[bend left] node[above] { \{\},$x>T$ } (q)
             (p2) edge node[above]  { \{$!rA_{out}$\} } (p3)
             (p3) edge node[above]  { \{$?rA_{in}$\},x:=0} (p4)
             (p4) edge node[above]  { \{\},$x>T$ } (q)
             (p4) edge node[near start]  { \{$!wA_{in}$\}} (q1);
     \end{tikzpicture}
     \label{fig:TACA_mixed}
    }}
    \end{tabular}
\caption{Handshaking behavior of source, sink, and simple mixed nodes}
\label{fig:TACA_simple}
\end{figure}

Figure~\ref{fig:TACA_simple} shows the TACA for simple Reo nodes: a \emph{source} node with one output port $A_{out}$, a \emph{sink} node with one input port $A_{in}$, and a \emph{mixed} node with one input port $A_{in}$ and one output port $A_{out}$.
We use a set of action types $\calT_1=\{?, !\} \times \{w, r, mw\}$ to represent sending and receiving of \emph{write}, \emph{read}, and \emph{may\_write} messages, respectively. We also use a set of actions $\calT_2=\{b, u\}$ to define blocking and unblocking of node ports.
Thus, a set of action names derived from the set of port names $\{A_{in}, A_{out}\}$ and a set of admissible action types $\calT = \calT_1 \cup \calT_2$, is $\N = \N_{in} \cup \N_{out}$ where $$\N_{in} = \{?wA_{in}, !wA_{in}, ?rA_{in}, !rA_{in}, ?mwA_{in}, !mwA_{in}, bA_{in}, uA_{in}\}$$ is a set of actions observed on the node's input port, and $$\N_{out} = \{?wA_{out}, !wA_{out}, ?rA_{out}, !rA_{out}, ?mwA_{out}, !mwA_{out}, bA_{out}, uA_{out}\}$$ is the set of actions observed on its output port. If a node performs the same action on all its ports simultaneously, we write $\alpha A \,| \, \alpha \in \calT.$ For example, $bA$ and $uA$ stand for ``block/unblock all ports of node $A$'', respectively.

The handshaking behavior of a \emph{source} node is shown in Figure~\ref{fig:TACA_source}). The source node $A$ sends a \emph{write} message through its output port and waits for a reply for the time $x < T,$ where $T$ is a timeout large enough to guarantee that any message in the synchronous region of the circuit is propagated to the most remote (in terms of the communication delay) node and back. If the node does not receive a reply from the accepting party, it assumes that the latter is not ready to read and discards the request. If the answer is received (i.e., $!rA_{out}$), the node commits to transfer data by blocking its ports ($bA$). Finally, after the timeout expires, the node unblocks its ports ($uA$) and returns to the initial state.

The \emph{sink} node (see Figure~\ref{fig:TACA_sink}) is a passive node that waits for the \emph{write} or \emph{may\_write} messages. After such a message is received, it either confirms its readiness to accept ($?rA_{in}$) or ignores the message and returns to the initial state. The transition shown with the dashed line is not required for always accepting sink nodes. If the \emph{write} message is received and the node is ready to accept, it goes to the committed state. If the \emph{may\_write} was received and the node confirmed the intention to accept, it awaits for the confirmation to write ($!wA_{in}$), replies to it and only then commits.

The \emph{mixed} node exhibits the behavior which is the combination of the above. It accept the incoming messages and forwards the them to its neighbor though the output port ($?wA_{out}$ and $?mwA_{out}$), waits for the reply ($!rA_{out}$) and forwards it back through the input port ($?rA_{in}$). Similarly, if $!wA_{in}$ is received after $!mwA_{in}$, the mixed node forwards the \emph{write} request to the output port, waits for the acknowledgement, forwards it to the sender, and commits.

Note that it is important to acknowledge both \emph{may\_write} and \emph{write} messages to be able to propagate I/O information in the circuit: the \emph{read} message in response to a \emph{may\_write} message is just a confirmation that some accepting sink node exists in the circuit, yet the data transfer through a particular node may not happen due to the non-deterministic choices of internal Reo nodes. In contrast, the \emph{read} message in response to a \emph{write} message means that the exchanging parties agreed to transfer data.

Considering the presence of $may\_write$ messages followed by \emph{write} messages, $T$ can be roughly estimated as $C$ times the longest path in the (synchronous region of the) Reo circuit graph, where $C$ is a constant that depends on the number of branches for merge nodes that may receive \emph{may\_write} messages. We will address the issue of computing the lower bound on timeout delays in our future work.

The handshaking behavior of the replicate node with input port $A_{in}$ and output ports $A_{out1}$ and $A_{out2}$ is shown in Figure~\ref{fig:TACA_replicate}. Once a \emph{write} or \emph{may\_write} message is received, the node sends a \emph{may\_write} message to its both output ports and waits for the replies. If meanwhile the timeout expires, the node returns to its initial state. If both neighbors confirm their ability to read, the further processing depends on the status of the input port: if the \emph{write} message was received initially, the node $A$ knows that it is able to provide data to its output ports and thus sends \{$?wA_{out1}, ?wA_{out2}$\} to agree on the certain data exchange with them.
If both $!rA_{out1}$ and $!rA_{out2}$ are observed, $A$ forwards the reply back through its input port and commits ($bA_{in}$).
Alternatively, if $!mwA_{in}$ initially triggered the decision making cycle on $A$, it has to request for the confirmation of the flow ($?rA_{in}$), and if it is confirmed ($!wA_{in}$), proceed as before. The difference in the procedure for the triggering \emph{may\_write} message is shown with dashed lines.

\begin{figure}
 \centering
 \scalebox{0.7}{
    \begin{tikzpicture}
            \node[state] (q) {$s_0$};
            \node[state,right of=q] (q1) {};
            \node[state,right of=q1] (q2) {$x<T$};
            \node[state,below right of=q2] (q3a) {$x<T$};
            \node[state,above right of=q2] (q3b) {$x<T$};
            \node[state,below right of=q3b] (q3) {};
            \node[state,below right of=q3] (q4) {$x<T$};
            \node[state,below left of=q4] (q5) {};
            \node[state,left of=q5] (q6) {$x<T$};
            \node[state,below left of=q6] (q7a) {$x<T$};
            \node[state,above left of=q6] (q7b) {$x<T$};
            \node[state,below left of=q7b] (q7) {};
            \node[state,left of=q7] (q8) {};
            \node[state,above left of=q8] (commit) {$x<T$};
        \path[transition]
             (q)  edge node[above]  {\{$!wA_{in}$\}} (q1)
             (q)  edge[bend left, dashed] node[above]  {\{$!mwA_{in}$\}} (q1)
             (q1) edge node[above] {\{$?mwA_{out1},?mwA_{out2}$\},x:=0} (q2)
             (q2) edge[bend right] node[above] {\{\},$x>T$} (q)
             (q2) edge node {\{$!rA_{out1}$\}} (q3a)
             (q2) edge node {\{$!rA_{out2}$\}} (q3b)
             (q3a) edge node {\{$!rA_{out2}$\}} (q3)
             (q3b) edge node {\{$!rA_{out1}$\}} (q3)
             (q3a) edge node[above]  {\{\},$x>T$} (q)
             (q3b) edge[bend right] node[above]  {\{\},$x>T$} (q)
             (q2) edge node[above]  {\{$!rA_{out1},!rA_{out2}$\}} (q3)
             (q3) edge node[above]  {\{\}} (q5)
             (q3) edge[dashed] node {\{$?rA_{in}$\},x:=0} (q4)
             (q4) edge[dashed] node {\{$!wA_{in}$\}} (q5)
             (q4) edge[dashed] node[above] {\{\},$x>T$} (q)
             (q5) edge node[above] {\{$?wA_{out1},?wA_{out2}$\},x:=0} (q6)
             (q6) edge node {\{$!rA_{out1}$\}} (q7a)
             (q6) edge node {\{$!rA_{out2}$\}} (q7b)
             (q7a) edge node {\{$!rA_{out2}$\}} (q7)
             (q7b) edge node {\{$!rA_{out1}$\}} (q7)
             (q7a) edge[bend left]  node[above]  {\{\},$x>T$} (q)
             (q7b) edge node[above]  {\{\},$x>T$} (q)
             (q6) edge node[above]  {\{$!rA_{out1},!rA_{out2}$\}} (q7)
             (q7) edge node[above]  {\{$!rA_{in}$\}} (q8)
             (q8) edge node[above]  {\{$bA$\},x:=0} (commit)
             (commit) edge node[above] {\{$uA$\},$x>T$} (q)
            ;
     \end{tikzpicture}
  }
\caption{Replicate node}
\label{fig:TACA_replicate}
\end{figure}

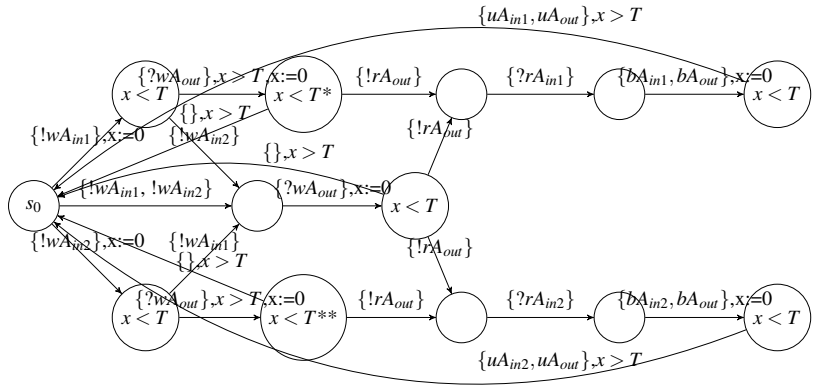
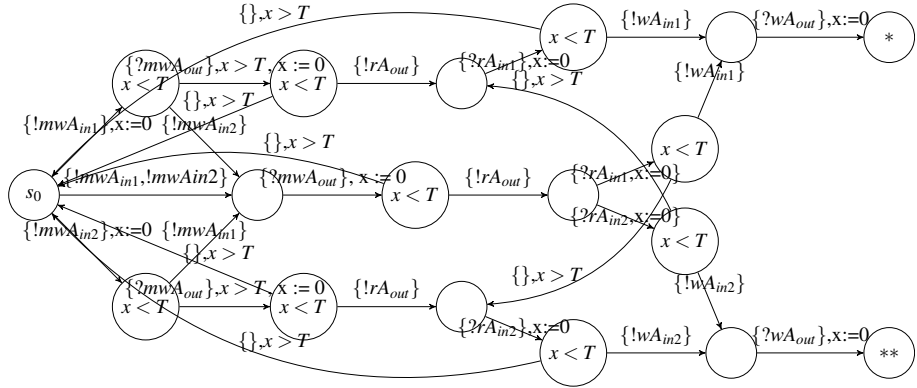
\begin{figure}
 \centering
    \begin{tabular}{c}
    \subfigure[\scriptsize{Merge node with both source ends ready to write}]{
    \scalebox{0.7}{
    \begin{tikzpicture}
            \node[state] (q) {$s_0$};
            \node[state,above right of=q] (q1a) {$x<T$};
            \node[state,below right of=q] (q1b) {$x<T$};
            \node[state,below right of=q1a] (q1) {};
            \node[state,right of=q1a] (q2a) {$x<T$*};
            \node[state,right of=q1b] (q2b) {$x<T$**};
            \node[state,right of=q1] (q2) {$x<T$};
            \node[state,right of=q2a] (q3a) {};
            \node[state,right of=q2b] (q3b) {};
            \node[state,right of=q3a] (q4a) {};
            \node[state,right of=q3b] (q4b) {};
            \node[state,right of=q4a] (commit1) {$x<T$};
            \node[state,right of=q4b] (commit2) {$x<T$};
        \path[transition]
             (q)  edge node[above]  { \{$!wA_{in1}$, $!wA_{in2}$\} } (q1)
             (q) edge node[above]  { \{$!wA_{in1}$\},x:=0} (q1a)
             (q) edge node[above]  { \{$!wA_{in2}$\},x:=0} (q1b)
             (q1a) edge node[above]  { \{$!wA_{in2}$\} } (q1)
             (q1b) edge node[above]  { \{$!wA_{in1}$\} } (q1)
             (q1a) edge node[above]  { \{$?wA_{out}$\},$x>T$,x:=0} (q2a)
             (q1b) edge node[above]  { \{$?wA_{out}$\},$x>T$,x:=0} (q2b)
             (q1)  edge node[above]  { \{$?wA_{out}$\},x:=0} (q2)
             (q2) edge[bend right=20] node[above,near start] { \{\},$x>T$ } (q)
             (q2a) edge node[above]  { \{$!rA_{out}$\}} (q3a)
             (q2b) edge node[above]  { \{$!rA_{out}$\}} (q3b)
             (q2a) edge node[above,near start]  { \{\},$x>T$} (q)
             (q2b) edge node[above,near start]  { \{\},$x>T$} (q)
             (q2)  edge node[above]  { \{$!rA_{out}$\}} (q3a)
             (q2)  edge node[above]  { \{$!rA_{out}$\}} (q3b)
             (q3a)  edge node[above]  { \{$?rA_{in1}$\}} (q4a)
             (q3b)  edge node[above]  { \{$?rA_{in2}$\}} (q4b)
             (q4a)  edge node[above]  { \{$bA_{in1},bA_{out}$\},x:=0} (commit1)
             (q4b)  edge node[above]  { \{$bA_{in2},bA_{out}$\},x:=0} (commit2)
             (commit1)  edge[bend right] node[near start, above]  { \{$uA_{in1},uA_{out}$\},$x>T$} (q)
             (commit2)  edge[bend left] node[near start, above]  { \{$uA_{in2},uA_{out}$\},$x>T$} (q)
             ;
     \end{tikzpicture}
     \label{fig:TACA_merge1}
     }}
     \\
     \subfigure[\scriptsize{Merge node with both source ends that may write}]{
     \scalebox{0.7}{
     \begin{tikzpicture}
            \node[state] (q) {$s_0$};
            \node[state,above right of=q] (q1a) {$x<T$};
            \node[state,below right of=q] (q1b) {$x<T$};
            \node[state,below right of=q1a] (q1) {};
            \node[state,right of=q1a] (q2a) {$x<T$};
            \node[state,right of=q1b] (q2b) {$x<T$};
            \node[state,right of=q1] (q2) {$x<T$};
            \node[state,right of=q2] (q3) {};
            \node[state,right of=q2a] (q3a) {};
            \node[state,right of=q2b] (q3b) {};
            \node[state,above of=q3] (q4a) {$x<T$};
            \node[state,below of=q3] (q4b) {$x<T$};
            \node[state,below right of=q4a] (q4aa) {$x<T$};
            \node[state,above right of=q4b] (q4bb) {$x<T$};
            \node[state,right of=q4a] (q5a) {};
            \node[state,right of=q4b] (q5b) {};
            \node[state,right of=q5a] (q6a) {$*$};
            \node[state,right of=q5b] (q6b) {$**$};
        \path[transition]
             (q) edge node[above]  { \{$!mwA_{in1}$,$!mwA{in2}$\} } (q1)
             (q) edge node[above]  { \{$!mwA_{in1}$\},x:=0} (q1a)
             (q) edge node[above]  { \{$!mwA_{in2}$\},x:=0} (q1b)
             (q1a) edge node[above]  { \{$!mwA_{in2}$\} } (q1)
             (q1b) edge node[above]  { \{$!mwA_{in1}$\} } (q1)
             (q1a) edge node[above]  { \{$?mwA_{out}$\},$x>T$, x := 0} (q2a)
             (q1b) edge node[above]  { \{$?mwA_{out}$\},$x>T$, x := 0} (q2b)
             (q1)  edge node[above]  { \{$?mwA_{out}$\}, x := 0} (q2)
             (q2) edge[bend right=20] node[above, near start] { \{\},$x>T$ } (q)
             (q2a)  edge node[above]  { \{$!rA_{out}$\}} (q3a)
             (q2b)  edge node[above]  { \{$!rA_{out}$\}} (q3b)
             (q2a) edge node[above, near start]  { \{\},$x>T$} (q)
             (q2b) edge node[above, near start]  { \{\},$x>T$} (q)
             (q2)  edge node[above]  { \{$!rA_{out}$\}} (q3)
             (q3)  edge node {\{$?rA_{in1}$,x:=0\}} (q4aa)
             (q3)  edge node {\{$?rA_{in2}$,x:=0\}} (q4bb)
             (q4aa)  edge node[above]  { \{$!wA_{in1}$\}} (q5a)
             (q4bb)  edge node[above]  { \{$!wA_{in2}$\}} (q5b)
             (q4aa)  edge[bend left] node[above, near end]  { \{\},$x>T$} (q3b)
             (q4bb)  edge[bend right] node[above, near end]  { \{\},$x>T$} (q3a)
             (q3a)  edge node {\{$?rA_{in1}$\},x:=0} (q4a)
             (q3b)  edge node {\{$?rA_{in2}$\},x:=0} (q4b)
             (q4a)  edge node[above]  { \{$!wA_{in1}$\}} (q5a)
             (q4b)  edge node[above]  { \{$!wA_{in2}$\}} (q5b)
             (q4a)  edge[bend right] node[above]  { \{\},$x>T$} (q)
             (q4b)  edge[bend left] node[above]  { \{\},$x>T$} (q)
             (q5a)  edge node[above]  { \{$?wA_{out}$\},x:=0} (q6a)
             (q5b)  edge node[above]  { \{$?wA_{out}$\},x:=0} (q6b);
     \end{tikzpicture}
     \label{fig:TACA_merge2}
     }}
     \end{tabular}
    \caption{Merge node}
    \label{fig:TACA_merge}
\end{figure}

The behavior of the \emph{merge} node with two input ports $A_{in1}$ and $A_{in2}$ and one output port $A_{out}$ depends on the type of the incoming messages: both \emph{write}, both \emph{may\_write}, or the combination of \emph{write} and \emph{may\_write}.
In the first case (see Figure~\ref{fig:TACA_merge1}), the merge node in its initial state waits for an incoming message on at least one of its input ports, $!wA_{in1}$, $!wA_{in2}$ or both, forwards the incoming \emph{write} message to the output port ($?wA_{out}$), and waits for the confirmation to read ($!rA_{out}$). If one input message is received, the node waits for the incoming message on the other input port.
If the second message does not arrive within the timeout, it means that the other writer is not ready to write and $A$ will proceed with the available request. If both input ports received a write request, the merge node chooses an incoming port to accept data from. Once the decision is made, the node $A$ sends the \emph{read} message to the selected input port ($?rA_{in1}$ or $?rA_{in2}$) and commits (\{$bA_{in1}, bA_{out}$\} or \{$bA_{in2}, bA_{out}$\}).

In Figure~\ref{fig:TACA_merge2}, $!mwA_{in1}$ and $!mwA_{in2}$ are initially received. This means that the sender tries to establish which transactions are enabled. Consequently, after propagating the initial \emph{may\_write} message to the output port ($?mwA_{out}$) and receiving $!rA_{out}$, the node $A$ needs to wait for the definitive confirmation to provide data from at least one of its input ends, $!wA_{in1}$ or $!wA_{in2}$. If such a message arrives, the node commits, otherwise, returns to the initial state. In the case when both input nodes are ready to write, $A$ non-deterministically chooses one of the branches and sends the \emph{read} message to it, either $?rA_{in1}$ or $?rA_{in2}$. If the selected input port receives the confirmation to write, the node forwards it to the output port and commits. However, the node that issued the initial \emph{may\_write} message may choose to fire a different transaction and the node $A$ will not receive a \emph{write} message. This should not exclude the transaction that involves the pending request on its second input port. Thus, on the timeout the node $A$ sends the \emph{read} message to the remaining source port. If the \emph{write} message is finally received ($?wA_{out}$), the node forwards it to its output port and processes further messages as in Figure~\ref{fig:TACA_merge1}. Otherwise, due to the external choices, the node does not participate in the data transfer at this cycle and returns to its initial state.
The combination of \emph{write} and \emph{may\_write} messages yield an automaton with the behavioral patterns shown in the above two cases.

The \emph{route} node with one input port $A_{in}$ and two output ports $A_{out1}$ and $A_{out2}$ receives a \emph{write} message $!wA_{in}$ or a \emph{may\_write} message $!mwA_{in}$, forwards it to its output ports and waits for the \emph{read} messages. Figure~\ref{fig:TACA_route} shows the case for the initial \emph{write} message, the \emph{may\_write} is processed in a similar way and is omitted due to the lack of space. After at least one of the output ports confirms the ability to read ($!rA_{out1}$ and/or $!rA_{out2}$), the node $A$ non-deterministically chooses among the available options and confirms the intention to write ($?wA_{out1}$ or $?wA_{out2}$). If the confirmation is received, the node proceeds by sending to the input port $?rA_{in}$ and commits (\{$bA_{in}, bA_{out1}$\} or \{$bA_{in}, bA_{out1}$\}). Alternatively, on the expiration of the timeout, the route node tries to confirm the intention to write to another enabled output port if such an option is available.

For the initial \emph{may\_write} message the mechanism is similar, but before the node decides to issue a definite write message, it needs to confirm the ability to read $!rA_{in}$, receive the definite \emph{write} message $?wA_{in},$ make a choice between $?wA_{out1}$ and $?wA_{out2},$ make sure that the port it had chosen acknowledged the ability to read (or try other options alternatively), send another confirmation to read to its input port ($!rA_{in}$) and commit.
\begin{figure}
 \centering
    \scalebox{0.7}{
    \begin{tikzpicture}
            \node[state] (q) {$s_0$};
            \node[state,right of=q] (q1) {};
            \node[state,right of=q1] (q2) {$x<T$};
            \node[state,above right of=q2] (q3a) {$x<T$};
            \node[state,below right of=q2] (q3b) {$x<T$};
            \node[state,below right of=q3a] (q3) {};
            \node[state,right of=q3a] (q4a) {};
            \node[state,right of=q3b] (q4b) {};
            \node[state,above right of=q4a] (q5a) {$x<T$};
            \node[state,below right of=q4b] (q5b) {$x<T$};
            \node[state,right of=q4a] (q5ab) {$x<T$};
            \node[state,right of=q4b] (q5ba) {$x<T$};
            \node[state,right of=q5ab] (q6a) {};
            \node[state,right of=q5ba] (q6b) {};
            \node[state,right of=q6a] (q7a) {};
            \node[state,right of=q6b] (q7b) {};
            \node[state,right of=q7a] (commit1) {$x<T$};
            \node[state,right of=q7b] (commit2) {$x<T$};
        \path[transition]
             (q)  edge node[above]  { \{$!wA_{in}$\} } (q1)
             (q1) edge node[above]  { \{$?mwA_{out1}$,$?mwA_{out2}$\},x:=0} (q2)
             (q2) edge[bend right=20] node[above] { \{\},$x>T$} (q)
             (q2) edge node[above]  { \{$!rA_{out1}$\} } (q3a)
             (q2) edge node[above]  { \{$!rA_{out2}$\} } (q3b)
             (q3a) edge node[above]  { \{$!rA_{out2}$\} } (q3)
             (q3b) edge node[above]  { \{$!rA_{out1}$\} } (q3)
             (q2) edge node[above]  { \{$!rA_{out1}$,$!rA_{out2}$\} } (q3)
             (q3a) edge node[above]  { \{\},$x>T$} (q4a)
             (q3b) edge node[above]  { \{\},$x>T$} (q4b)
             (q4a) edge node[above] { \{$?wA_{out1}$\},x:=0} (q5a)
             (q4b) edge node[above] { \{$?wA_{out2}$\},x:=0} (q5b)
             (q3) edge node { \{$?wA_{out1}$\},x:=0} (q5ab)
             (q3) edge node { \{$?wA_{out2}$\},x:=0} (q5ba)
             (q5ab) edge node[near start]  { \{\},$x>T$} (q4b)
             (q5ba) edge node[near start]  { \{\},$x>T$} (q4a)
             (q5ab) edge node[above] { \{$!rA_{out1}$\}} (q6a)
             (q5ba) edge node[above] { \{$!rA_{out2}$\}} (q6b)
             (q5a) edge node[above]  { \{$!rA_{out1}$\}} (q6a)
             (q5b) edge node[above]  { \{$!rA_{out2}$\}} (q6b)
             (q5a) edge[bend right=15] node[above]  { \{\},$x>T$} (q)
             (q5b) edge[bend left=15] node[above]  { \{\},$x>T$} (q)
             (q6a) edge node[above]  { \{$?rA_{in}$\}} (q7a)
             (q6b) edge node[above]  { \{$?rA_{in}$\}} (q7b)
             (q7a) edge node[above] {\{$bA_{in},bA_{out1}$\},x:=0} (commit1)
             (q7b) edge node[above] {\{$bA_{in},bA_{out2}$\},x:=0} (commit2)
             (commit1) edge[bend right=34] node[above, near start]  { \{$uA$\},$x>T$ } (q)
             (commit2) edge[bend left=34] node[above, near start]  { \{$uA$\},$x>T$ } (q);
     \end{tikzpicture}
     }
  \caption{Route node}
  \label{fig:TACA_route}
\end{figure}

For the \emph{join} node $A$ to commit, both its input ports $A_{in1}$ and $A_{in2}$ should receive \emph{write} (or \emph{may\_write} with the consequent confirmation to write) messages, and its sink port should be able to read. If only one input request is received within the timeout, it is discarded.
If both messages are \emph{write} messages, the node $A$ simply checks whether the sink end can read ($?wA_{out}$ followed by $!rA_{out}$), confirms the possibility of the flow to both senders (\{$!rA_{in1},!rA_{in2}$\}) and commits.

\begin{figure}
 \centering
    \scalebox{0.7}{
    \begin{tikzpicture}
            \node[state] (q) {$s_0$};
            \node[state,above right of=q] (q1a) {$x<T$};
            \node[state,below right of=q] (q1b) {$x<T$};
            \node[state,below right of=q1a] (q1) {};
            \node[state,right of=q1] (q2) {$x<T$};
            \node[state,right of=q2] (q3) {};
            \node[state,right of=q3] (q4) {$x<T$};
            \node[state,right of=q4] (q5) {};
            \node[state,below left of=q5] (q6) {$x<T$};
            \node[state,left of=q6] (q7) {};
            \node[state,left of=q7] (q8) {};
            \node[state,left of=q8] (commit) {$x<T$};
        \path[transition]
             (q)  edge node[above]  { \{$!mwA_{in1},!wA_{in2}$\} } (q1)
             (q) edge node[above]  { \{$!mwA_{in1}$\},x:=0} (q1a)
             (q) edge node[above]  { \{$!wA_{in2}$\},x:=0} (q1b)
             (q1a) edge node[above]  { \{$!wA_{in2}$\} } (q1)
             (q1b) edge node[above]  { \{$!mwA_{in1}$\} } (q1)
             (q1a) edge[bend right] node[above]  { \{\},$x>T$} (q)
             (q1b) edge[bend left] node[above]  { \{\},$x>T$} (q)
             (q1)  edge node[above]  { \{$?mwA_{out}$\},x:=0} (q2)
             (q2) edge[bend right] node[above, near start]  {\{\},$x>T$} (q)
             (q2)  edge node[above]  { \{$!rA_{out}$\}} (q3)
             (q3)  edge node[above]  { \{$?rA_{in1}$\},x:=0} (q4)
             (q4) edge[bend right] node[above, near start]  {\{\},$x>T$} (q)
             (q4)  edge node[above]  { \{$!wA_{in1}$\}} (q5)
             (q5)  edge node[above]  { \{$?wA_{out}$\}} (q6)
             (q6) edge node[above, near start]  {\{\},$x>T$} (q)
             (q6)  edge node[above]  { \{$!rA_{out}$\}} (q7)
             (q7)  edge node[above]  { \{$?rA_{in1},?rA_{in2}$\}} (q8)
             (q8)  edge node[above]  { \{$bA$\},x:=0} (commit)
             (commit)  edge[near start] node[above]  {\{$uA$\},$x>T$} (q)
             ;
     \end{tikzpicture}
     }
    \caption{Join node}
    \label{fig:TACA_join}
\end{figure}
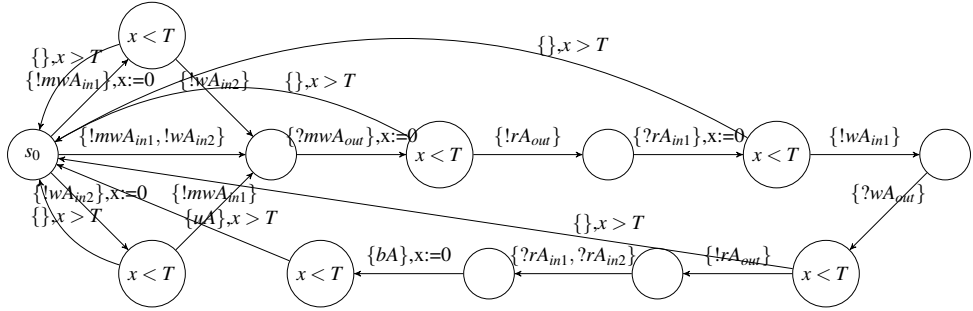

Figure~\ref{fig:TACA_join} shows a less obvious case with one \emph{may\_write} input message ($!mwA_{in1}$) and one \emph{write} message ($!wA_{in2}$).
The join node cannot guarantee the flow to its sink and thus senses whether the sink end can read with an uncertain \emph{may\_write} request ($?mwA_{out}$). If $!rA_{out}$ follows, the node should first request the uncertain input port to confirm the intention to write ($?rA_{in1}$). If so, the node behaves as in the previous case: it sends the \emph{write} messages to its sink end, waits for the confirmation, forwards the confirmation to both input ports (\{$?rA_{in1},?rA_{in2}$\}) and commits.

Data flow in a Reo circuit is controlled not only by its nodes, but also by its channels that may perform operations on the data they receive.
What is a channel and how do we implement them? Essentially, each channel is an abstraction for a set of hops in a computer network to and from a component that implements the channel's behavioral logic; its source port is known to data suppliers while its sink port is known to data consumers. From the viewpoint of handshaking protocol, Reo channels are communication links to exchange messages between adjacent nodes (e.g., {\Sync} channel). However, channels behave also like nodes that determine which transitions are enabled (e.g., {\SyncDrain} and {\AsyncDrain}).

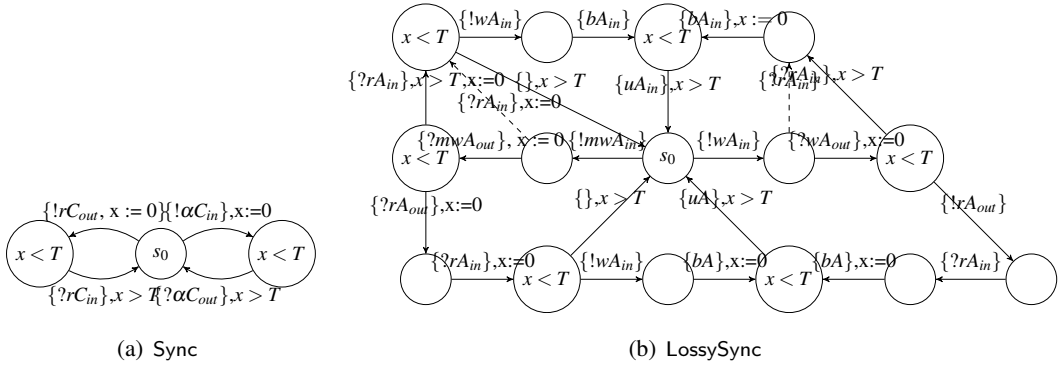
\begin{figure}
 \centering
     \subfigure[\scriptsize{{\Sync}}]{
     \scalebox{0.7}{
     \label{fig:TACA_sync}
     \begin{tikzpicture}
             \node[state] (q) {$s_0$};
             \node[state,right of=q] (q1) {$x<T$};
             \node[state,left of=q] (p1) {$x<T$};
         \path[transition]
              (q)  edge[bend left] node[above]  { \{$!\alpha C_{in}$\},x:=0} (q1)
              (q1) edge[bend left] node[below] { \{$?\alpha C_{out}$\},$x>T$} (q)
              (q) edge[bend right] node[above]  { \{$!rC_{out}$, x := 0\} } (p1)
              (p1) edge[bend right] node[below]  { \{$?rC_{in}$\},$x>T$ } (q);
      \end{tikzpicture}
    }}
    \subfigure[\scriptsize{{\LossySync}}]{
    \scalebox{0.7}{
    \label{fig:TACA_lossysync}
    \begin{tikzpicture}
            \node[state] (q) {$s_0$};
            \node[state,right of=q] (q1) {};
            \node[state,right of=q1] (q2) {$x<T$};
            \node[state,below of=q2] (q3) {};
            \node[state,right of=q3] (q4) {};
            \node[state,below of=q1] (commit) {$x<T$};
            \node[state,left of=q] (p1) {};
            \node[state,left of=p1] (p2) {$x<T$};
            \node[state,below of=p2] (p3) {};
            \node[state,below of=p1] (p4) {$x<T$};
            \node[state,below of=q] (p5) {};
            \node[state, above of=q] (commit1) {$x<T$};
            \node[state, above of=p2] (lost2) {$x<T$};
            \node[state, above of=p1] (lost1) {};
            \node[state, above of=q1] (q1a) {};
        \path[transition]
             (q)  edge node[above]  { \{$!wA_{in}$\} } (q1)
             (q1) edge node[above]  { \{$?wA_{out}$\},x:=0} (q2)
             (q2) edge node[near end] { \{$?rA_{in}$\},$x>T$ } (q1a)
             (q1) edge[dashed] node[above] { \{$?rA_{in}$\}} (q1a)
             (q1a) edge node[above] { \{$bA_{in}$\},$x := 0$ } (commit1)
             (q2) edge node[above]  { \{$!rA_{out}$\} } (q4)
             (q4) edge node[above]  { \{$?rA_{in}$\}} (q3)
             (q3) edge node[above]  { \{$bA$\},x:=0} (commit)
             (commit) edge node[above]  { \{$uA$\},$x>T$ } (q)
             (commit1) edge node[above]  { \{$uA_{in}$\},$x>T$ } (q)
             (q)  edge node[above]  { \{$!mwA_{in}$\} } (p1)
             (p1) edge node[above]  { \{$?mwA_{out}$\}, x := 0} (p2)
             (p2) edge node[above] { \{$?rA_{out}$\},x:=0} (p3)
             (p2) edge node[above] { \{$?rA_{in}$\},$x>T$,x:=0} (lost2)
             (p1) edge[dashed] node[above, near start] { \{$?rA_{in}$\},x:=0} (lost2)
             (lost2) edge node[above] { \{$!wA_{in}$\}} (lost1)
             (lost2) edge node[above] { \{\},$x>T$} (q)
             (lost1) edge node[above]  { \{$bA_{in}$\} } (commit1)
             (p3) edge node[above]  { \{$?rA_{in}$\},x:=0} (p4)
             (p4) edge node[above]  { \{\},$x>T$ } (q)
             (p4) edge node[above]  { \{$!wA_{in}$\}} (p5)
             (p5) edge node[above]  { \{$bA$\},x:=0} (commit);
     \end{tikzpicture}
     }}
    \caption{Handshaking behavior of {\Sync} and {\LossySync} channels}
\end{figure}

The handshaking behavior of a {\Sync} channel $c$ with communication delay $t$, source port $C_{in}$ and sink port $C_{out}$ can be modeled as shown in Figure~\ref{fig:TACA_sync}. It accepts a \emph{write} or a \emph{may\_write} message on its source end ($!\alpha C_{in}$ where $\alpha \in \{m, mw\}$) and notifies its sync end ($?\alpha C_{out}$). Similarly, it accepts a \emph{read} message on its sink end ($!rC_{out}$) and transfers it to its source end ($?rC_{in}$).

The behavior of the {\LossySync} channel as a transition link is analogous. However, its ability to lose data either when its output node is not ready to accept (context-dependent {\LossySync}) or non-deterministically (context-independent {\LossySync}) should be modeled explicitly. In the first case, the decision to accept data on the source end of the {\LossySync} channel despite the fact that its sink end is unable to read can be modeled as shown in Figure~\ref{fig:TACA_lossysync}. This figure shows the handshaking behavior of an always accepting mixed node that commits to a transaction that involves only its input port after the timeout of waiting for a \emph{read} message from its output port expires. Thus, a context-dependent {\LossySync} channel can be represented as a {\Sync} channel joint to such a node.

The context-independent {\LossySync} behaves similarly, but it may accept data without notifying its sink port, as shown by two dashed transitions $?bA_{in}$ in Figure~\ref{fig:TACA_lossysync}, in right and left branches.
\begin{figure}
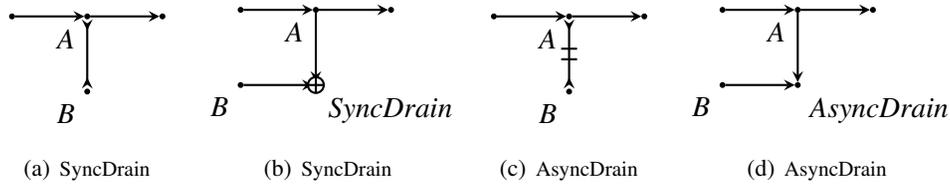

\centering
       \subfigure[\scriptsize{SyncDrain}]{
       \scalebox{1.0}{
            \tikz{
                 \node[point] at (-1, 1) (X) {};
                 \node[point,label=below left:$A$] at (0, 1) (A) {};
                 \node[point,label] at (1, 1) (Y) {};
                 \node[point,label=below left:$B$] at (0, 0) (B) {};
                 \draw[sync]  (X) to node {} (A);
                 \draw[sync]  (A) to node {} (Y);
                 \draw[syncdrain] (A) to node {} (B);
            }
       }}
       \subfigure[\scriptsize{SyncDrain}]{
       \scalebox{1.0}{
            \tikz{
                 \node[point] at (-1, 1) (X) {};
                 \node[point,label=below left:$A$] at (0, 1) (A) {};
                 \node[point,label] at (1, 1) (Y) {};
                 \node[point,label=below right:$SyncDrain$] at (0, 0) (SyncDrain) {};
                 \node[point,label=below left:$B$] at (-1, 0) (B) {};
                 \draw[sync]  (X) to node {} (A);
                 \draw[sync]  (A) to node {} (Y);
                 \draw[sync] (A) to node {} (SyncDrain);
                 \draw[sync] (B) to node {} (SyncDrain);
                 \draw [-, thick, fill=white] (0,0) circle (3pt);
                 \draw [-, thick](-0.1, 0) -- (0.1,0);
                 \draw [-, thick](0, -0.1) -- (0,0.1);
            }
        }}
        \subfigure[\scriptsize{AsyncDrain}]{
        \scalebox{1.0}{
           \tikz{
                 \node[point] at (-1, 1) (X) {};
                 \node[point,label=below left:$A$] at (0, 1) (A) {};
                 \node[point,label] at (1, 1) (Y) {};
                 \node[point,label=below left:$B$] at (0, 0) (B) {};
                 \draw[sync]  (X) to node {} (A);
                 \draw[sync]  (A) to node {} (Y);
                 \draw[asyncdrain] (A) to node {} (B);
            }
        }}
        \subfigure[\scriptsize{AsyncDrain}]{
        \scalebox{1.0}{
            \tikz{
                 \node[point] at (-1, 1) (X) {};
                 \node[point,label=below left:$A$] at (0, 1) (A) {};
                 \node[point,label] at (1, 1) (Y) {};
                 \node[point,label=below right:$AsyncDrain$] at (0, 0) (AsyncDrain) {};
                 \node[point,label=below left:$B$] at (-1, 0) (B) {};
                 \draw[sync]  (X) to node {} (A);
                 \draw[sync]  (A) to node {} (Y);
                 \draw[sync] (A) to node {} (AsyncDrain);
                 \draw[sync] (B) to node {} (AsyncDrain);
            }
        }}
  \caption{Modeling handshaking behavior of {\SyncDrain} and {\AsyncDrain} channels}
\label{fig:TACA_drains}
\end{figure}

Alternatively, a non-deterministic {\LossySync} can be modeled with the help of a router node that either chooses to pass data or reroute it to an unobservable always accepting output node $\tau$ (trash bin) where this data item is destroyed.
In the same fashion, the behavior of {\SyncDrain} and {\AsyncDrain} channels can be modeled with the help of auxiliary join and merge nodes as shown in Figure~\ref{fig:TACA_drains}.

%% file: coordination.tex
\section{Handshaking protocol correctness}
\label{sect:proof}

In this section, we outline the sketch of a proof that the presented handshaking TACA provides correct implementation for Reo.
To be able to show this formally, we need to define the notion of correct implementation.

\begin{definition}[Observable trace]
  \label{def:correctness}
  For a synchronous Reo circuit with a set of nodes $\calP$, let $\A = (S, \mathcal{C}, \N, \rightarrow_{\A}, s_0, ic),$ $\N = \calP \times \calT_{\A}$ be its handshaking TACA with port blocking, unblocking and auxiliary actions, $\{b,u\} \subset \calT_{\A}$.
  We say that $\longtrace{s}{N_1 \cdot N_2 \cdot ... \cdot N_n}{s'}$ is an \emph{observable} trace in $\A$ if it is an untimed trace in $\texttt{hide}(\A, \N \setminus \M)$ where $\M = \calP \times \calT_{\mathcal{B}},$ $\calT_{\mathcal{B}} = \{b, u\}.$
\end{definition}

\begin{definition}[Correct implementation]
  \label{def:correctness}
  For a synchronous Reo circuit with a set of nodes $\calP$, let $\mathcal{B} = (Q, \M, \rightarrow_{\mathcal{B}}, q_0)\, | \,  \M = \calP \times \calT_{\mathcal{B}}, \, \calT_{\mathcal{B}} = \{b, u\}$ be its ACA, and $\A = (S, \mathcal{C}, \N, \rightarrow_{\A}, s_0, ic) \, | \, \N = \calP \times \calT_{\A}$ where $\calT_{\A} \cap \calT_{\mathcal{B}} = \{b, u\}$ be its handshaking TACA.  The Reo handshaking protocol defined by $\A$ is a correct implementation of Reo iff there exists a mapping $\Theta: Q \rightarrow S$ such that
  \begin{itemize}
    \item for any $\transition{q}{M}{q'}$ in $\mathcal{B}$ there exists an observable trace $\longtrace{s}{N_1 \cdot N_2 \cdot ... \cdot N_n}{s'}$ in $\A$ such that $M = N_1 \cup N_2 \cup ... \cup N_n,$
    \item for any observable trace $\longtrace{s}{N_1 \cdot N_2 \cdot ... \cdot N_n}{s'}$ in $\A$, there exists $\transition{q}{M}{q'}$ in $\mathcal{B}$, $M = N_1 \cup N_2 \cup ... \cup N_n.$
  \end{itemize}
\end{definition}

Intuitively, this definition requires the handshaking protocol to block and unblock only those sets of ports that appear in synchronization constraints of port blocking ACA. Since auxiliary actions may be required in the implementation, blocking and unblocking of all involved ports does not need to be simultaneous, it is sufficient that for each ACA transition to have a state in TACA with all necessary ports blocked and later unblock these ports.

To show the correctness of our protocol, we should (i) show that TACA provide correct implementation of port blocking ACA for each basic Reo node and channel as defined in~\cite{KCA10}, (ii) using product operator, define TACA for a composed circuit by synchronizing message exchange actions on shared ports and show that such a TACA is a correct implementation for the composed circuit.


For each Reo node and channel, we hide all handshaking messages in the time-abstracted version of TACA and show that the obtained automaton is weakly bisimilar to the port-blocking ACA for this node or channel (or equivalently, the initial TACA is the action refinement of  the port blocking ACA). Compositionality of this relation helps to extend the proof to any Reo circuit. For a circuit with a set of nodes $\calP$, its handshaking behavior is given by
$$\A = (S, \mathcal{C}, \N, \rightarrow_{\A}, s_0, ic), \, \N = \calP \times \calT_{\A}, \, \calT_{\A} = \{b, u, ?w, !w, ?mw, !mw, ?r, !r\}.$$ It can be checked that for all basic Reo channels and nodes,  $\mathcal{B} \preccurlyeq \A$ with a set of hidden actions $K = \{\alpha \cdot X, \, \alpha \in \calT_{\A} \setminus \calT_{\mathcal{B}}, X \in \calP\}$ and an empty set of renamings $R=\emptyset.$
Since for each $\transition{q}{M}{q'}$ in $\mathcal{B}$ there exists $\trace{s}{M}{s'}$ in $\texttt{hide}(\A, \N \setminus \M),$ $\A$ is a correct implementation of $\mathcal{B}.$

To compose the TACA for handshaking, let us define a synchronization function
$\gamma : (\gamma_1, \gamma_2), \, \gamma_1: \N \rightarrow \N_1, \gamma_2: \N \rightarrow \N_2$ as follows.
For any two joint ports $A_{out}$ and $B_{in}$ in a Reo circuit,
    \begin{equation}
    \begin{array}{c}
       \gamma_1(?wA_{out}B_{in}) = ?wA_{out}, \quad \gamma_2(?wA_{out}B_{in}) = !wB_{in},\\
       \gamma_1(?mwA_{out}B_{in}) = ?mwA_{out}, \quad  \gamma_2(?mwA_{out}B_{in}) = !mwB_{in},\\
       \gamma_1(!rA_{out}rB_{in}) = !rA_{out}, \quad \gamma_2(!rA_{out}rB_{in}) = ?rB_{in}, \\
       \gamma_1(bA_{out}B_{in}) = bA_{out}, \quad \gamma_2(bA_{out}B_{in}) = bB_{in},\\
       \gamma_1(uA_{out}B_{in}) = uA_{out}, \quad \gamma_2(uA_{out}B_{in}) = uB_{in}.
    \end{array}
    \label{eq:taca-product}
    \end{equation}

The definition of the implementation correctness of the handshaking protocol in the form of TACA is a weaker requirement than the action refinement and our synchronizing function enables an automaton for only one instance of suitable implementations.
The TACA product without synchronizing blocking and unblocking actions in lines 4 and 5 of (\ref{eq:taca-product}) will also yield a correct implementation for the corresponding port blocking ACA, but not the action refinement. It also generates a significantly larger automaton that is harder to deal with formally.
However, in the distributed environment with communication delays it may be difficult to perform simultaneous port blocking and unblocking on remote nodes to signal that they are ready to transfer data. In practice, we are only interested in the existence of a state in which all ports involved into a firing transition are blocked. In the TACA product without synchronizing blocking and unblocking actions, Reo ports that are ready for firing transitions are blocked in any order. For example, in the case of a node with two ports, $bA_{out}$ and $bB_{in}$, their interleaving $bA_{out} || bB_{in} = bA_{out}.bB_{in} + bB_{in}.bA_{out} + bA_{out}|bB_{in}$ gives us traces with the same set $\{bA_{out}, bB_{in}\}$ of performed actions, which conforms to our definition of the correct implementation. 

%% file: discussion.tex
\section{Related work}
\label{sect:discussion}
Proen\c{c}a et al.~\cite{PCV+12} identifies five implementation approaches for Reo and offers their own framework, called Dreams.
The most straightforward approach is a \emph{speculative} approach: data is sent through the channels and rolled
back when an inconsistency arises. This approach has never been implemented. The \emph{automata-based} approach~\cite{MLA08} relies on CA semantics and pre-computed behavior at compile time. Implementations based on connector \emph{coloring}~\cite{MLA08}
compute all solutions for the behavior of each round and keep them in a routing table. Existing \emph{search-based} implementations rely on structural operational semantics and are implemented in Alloy~\cite{KSA+08} and Maude~\cite{MSA06}. The \emph{constraint-based} approach~\cite{CPL+11} applies SAT solving techniques to search for single solutions on each round.

Dreams~\cite{PCV+12} is the first working implementation of a distributed engine for Reo. It is based on a \emph{commit} and \emph{send} message exchange and has common traits with our protocol. However, our protocol is a form of a speculative approach that does not require rollbacks and unnecessary data propagation. We provide a theoretical model for distributed Reo implementation with desired characteristics~\cite{PCV+12}:
\begin{itemize}
  \item \emph{Decoupling}: each node can be deployed on a separate machine, transitions in synchronous regions fire independently, and handshaking message exchange occurs within the boundaries of synchronous regions.
  \item \emph{Scalability}: similarly to the Dreams, no global consensus is required. The behavior is computed per step, avoiding not only the state space explosion typical for the centralized implementation, but also the need for each node to remember the states of other nodes in the same synchronous region. In contrast to the Dreams, our implementation resolves non-determinism locally by each node or channel with inherent non-deterministic behavior. This is a more scalable approach because it does not use routing tables, the decision which transition to fire is taken in the distributed manner.
  \item \emph{Reconfiguration}:
    runtime reconfiguration of a connector will not require any global or regional changes.
    Once a node detects a change in its environment, it only needs to adjust its own behavior. For example, if a new channel is connected to an input port of a simple node, this node becomes a merge node and should behave accordingly. Formal protocol adjustment rules to enable reconfiguration will be developed in our future work.
  \item \emph{Fault tolerance}:
    with the exception of failures of committed nodes, our protocol is fault tolerant. The timeouts as opposed to the permanent locks in Dreams guarantee that a failure is treated as inability of a node to process data. Even in the case of a committed node failure, only current transaction is affected, at the next step (after timeout expires on each committed state) the remaining circuit resumes its normal operation.
\end{itemize}

In some applications of Reo such as coordination of processes in a multi-core execution environment, delays are negligible. Arriving requests are propagated through the network instantly and can be modeled using action synchronization. In~\cite{JHA13}, a Reo-to-C compiler which generates partially-distributed implementations for shared-memory platforms is presented. An optimization technique to improve the scalability of the generated code is later introduced~\cite{JHA14}. In~\cite{JA13a} and \cite{JSA14}, the formal details of an implementation with partial-distribution based on synchronous regions are developed. An early Reo-to-Java compiler, which generates completely centralized implementations is described in~\cite{JA13b} and used for service orchestration in~\cite{JSS+12}.

The approach proposed in this paper can be used as foundation for distributed implementation of other coordination approaches for agent- and component-based systems where global consensus of synchronous entities is required. As opposed to commits and rollbacks typically employed to implement transactional processes~\cite{BBM+07}, sensing messages like our \emph{may\_write} messages can be employed to elicit the states of remote entities before taking decisions locally.

%% file: conclusions.tex
\section{Conclusions and Future Work}
\label{sect:conclusions}
In this paper, we proposed a theoretical model for distributed implementation of Reo. We extended the definition of ACA with the notion of time, used this model to define what a correct implementation for Reo is, and described a handshaking protocol that complies with our definition of correct implementation. Our approach exhibits properties implied by Reo but not enforced in the previous implementations: the resolution of non-deterministic choices does not require centralized decisions, and the timeout mechanism ensures that the network does not lock due to a failure of a remote node or a channel. Consequently, this execution model is more suitable for real-time monitoring, QoS estimation, failure detection and reconfiguration.

Due to the space limitations we could not discuss some relevant aspects of the approach, most notably, computation of timeouts. We plan to extend the protocol to handle data constraints and implement a service that deploys and executes Reo circuits based on the presented model. The approach used in this paper can be applied to describe handshaking behavior of Reo nodes and channels assuming other propagation strategies: backward and two-side propagation.